\documentclass[journal=jpcbfk,articletitle=true,manuscript=article]{achemso}
  
\usepackage{graphicx}
\usepackage{amsmath}
\usepackage{amssymb}
\usepackage{dcolumn}
\usepackage{color}
\usepackage{natbib}
\usepackage{ctable}
\usepackage[sort&compress]{natbib}

\title{Thermodynamics of Hydration from the Perspective of the Molecular Quasi-Chemical Theory of Solutions}
\author{Dilipkumar N.\ Asthagiri}\email{dna6@rice.edu}
\affiliation{Department of Chemical and Biomolecular Engineering, Rice University, Houston, TX 77005, USA}
\author{Michael E. Paulaitis}\email{michaelp@jhmi.edu}
\affiliation{Center for Nanomedicine, Johns Hopkins School of Medicine, Baltimore, MD 21231}
\author{Lawrence R. Pratt}\email{lpratt@tulane.edu}
\affiliation{Department of Chemical and Biomolecular Engineering, Tulane University, New Orleans, LA 70118}

\begin{document}

\begin{abstract}
The quasi-chemical organization of the potential distribution theorem --- molecular quasi-chemical theory (QCT) --- enables practical calculations and also provides a conceptual framework for molecular hydration phenomena. QCT can be viewed from multiple perspectives: (a) As a way to regularize an  ill-conditioned statistical thermodynamic problem;  (b) As an introduction of and emphasis on the neighborship characteristics of a solute of interest; (c) Or as a way to include accurate electronic structure descriptions of near-neighbor interactions in defensible statistical thermodynamics by clearly defining neighborship clusters. The theory has been applied to solutes of a wide range of chemical complexity, ranging from ions that  interact  with water with both long-ranged and chemically intricate short-ranged interactions, to solutes that interact with water solely through traditional van~der~Waals interations, and including water itself. The solutes range in variety from monoatomic ions to chemically heterogeneous macromolecules. A notable feature of QCT is that in applying the theory to this range of solutes, the theory itself provides guidance on the necessary approximations and simplifications that can facilitate the calculations. In this Perspective, we develop these ideas and document them with examples that reveal the insights that can be extracted using the QCT formulation. 
\end{abstract}


\begin{tocentry}
\begin{center}
        \includegraphics[scale=1]{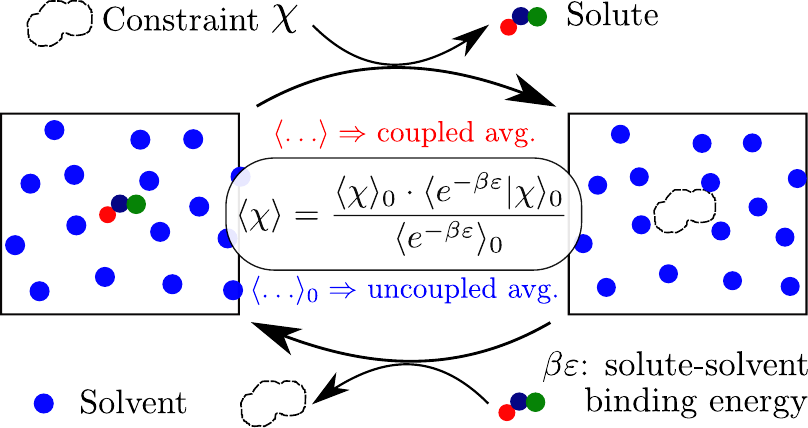}
\end{center}
\end{tocentry}

\section{Introduction}
The dissolution of a solute in a solvent or the precipitation of a
solute out of a solution are central to chemical transformations in
liquid solutions. For water as the solvent and the solutes being those
of interest in biology, these processes of dissolution (disaggregation)
or precipitation (aggregation) are central to life itself.  

A key challenge in physical chemistry has been to model, with the intent to predict, how solutes behave in a solution. In this quest, the excess chemical potential of Gibbs --- that part of the chemical potential that arises due to intermolecular interactions --- is
of foundational importance \cite{lrp:apc02,lrp:book}. With the growth of computer simulations, and attendant alchemical free energy calculations, important progress has been achieved in calculating and documenting the excess chemical potential (and its derivatives) for important categories of small molecules.  Over a decade ago, viewing the state of the field then, we noted ``\emph{the lack of a revealing theoretical model often means that simulation results are not as informative as they might be}''\cite{lrp:cpms}. This observation remains true even today,  despite the computational progress  of the intervening years.  An important feature of this ``\emph{\ldots not as informative as they might be}'' judgement is that molecular theory is a necessary feature of a proved molecular mechanism of hydration phenomena.  Molecular quasi-chemical theory (QCT) is proposed to supply the missing theory feature \cite{asthagiri:cpl10}.

QCT has provided important new insights into the phenomena of hydration of important classes of solutes.  In this Perspective, we highlight these developments and focus on hydration phenomena of 
current interest. QCT, being generally  developed statistical thermodynamics, of course naturally applies to all solutes, 
solvents, and solution phases. 

\subsection{First Steps in QCT Development}
The first steps in the development of QCT responded to the insistence that the 
statistical thermodynamic theory respect the results of electronic structure calculations on chemically relevant ions in water, {specifically} in the context of \emph{actinide molecular science} \cite{Martin:jpca98}.  A first application 
was to Li$^+$(aq) \cite{rempe:lijacs} because it was expected to be simpler than
other choices, and because of available influential 
neutron 
diffraction 
experiments on that system \cite{friedman1985hydration,howell1996hydration}.  Though
technically simple, QCT results  for Li$^+$(aq) disagreed with the 
experiments that motivated the calculations, 
specifically with respect to the most probable 
coordination of the Li$^+$(aq).  Furthermore,
there was no ready improvement of the theory that 
{could} resolve the disagreement.  The experiments were
refined over several years, and eventually the 
disagreement vanished  
\cite{rempe:bc06,mason2015neutron}, an exemplary instance of the scientific method. The moral of this vignette was that the concepts underlying 
QCT --- that chemically specific near-neighbor interactions
can be dominating, and that a defined inner-shell of a solute can be studied and characterized specifically --- provided robust tools in assisting experimental work.  

A lot was done immediately with that initial formulation
of QCT \cite{lrp:apc02}; examples include Refs.~\cite{Asthagiri:jacs04,Asthagiri:becpl,%
Asthagiri:hocpl,grabowski:jpca02}. With accumulating experience \cite{paliwal:jcp06}, it was soon noticed how a
more direct formulation of QCT provided a compelling statistical thermodynamic theory of liquid water 
itself \cite{shah:jcp07}.  That {formulation of} QCT was thoroughly tested 
as a direct description of water \cite{lrp:softcutoff,weber:jcp10a,weber:jcp10b};
indeed that QCT framework currently offers the pre-eminent
statistical thermodynamic theory of this most important
condensed material.  It is applicable to any of the force-field models that provide a physical basis for 
computations on liquid water, \emph{i.e.,} the model 
need not be adapted to the QCT.   The principal limitation, if at all, is merely that {it} is simplest to implement QCT calculations 
in the context of standard simulation calculations. That limitation could be relieved. The enabling simulations are often {still} simple enough and the QCT {approach} does indeed provide new information  \cite{shah:jcp07} 
that was not available from simulation alone.

We presented a unified QCT development \cite{asthagiri:cpl10} several years ago to document the singular
insights the theory brings to the problem of ion and small molecule hydration, especially in the context of biological selectivity of such molecules. (See also Ref.~\citenum{lrp:cpms}.) The perspective of the present discussion
is the important extension targeting QCT on macromolecular solutes of 
aqueous systems.  We note in passing the application
of QCT to polyethylene-oxide aqueous solutions, analyzing
the Flory-Huggins model descriptions of fluid phase 
separations of those systems \cite{chaudhari2014concentration}.
That work is distinct from the present focus in utilizing 
the small H$_2$O molecule  as a reporter species on the 
solution thermodynamics, and then building from the 
available Flory-Huggins model.

\subsection{Biomolecular Solutes}

Molecular quasi-chemical theory (QCT) was a break-through for the molecular theory of
liquid solutions \cite{lrp:apc02,lrp:book,lrp:cpms}, enabling
entirely new computational studies ranging from biological macromolecule hydration on the one hand to studies of water and aqueous ions using computationally demanding first principles calculations, on the other
hand.  In 2012,  in a first of its kind study, Weber and Asthagiri \cite{Weber:jctc12} tackled the
challenging problem of calculating the hydration free energy of a
protein, cytochrome C,  in an all-atom simulation. 
That work culminated over a decade of research in developing QCT, and
demonstrated that theoretical  refinements now make it possible \emph{to
calculate the hydration thermodynamics of bio-macromolecules at the same
resolution as for small molecules}, such as methane. Studies since then
have revealed fresh insights into the assumption of additivity of free
energy contributions \cite{paulaitis:corr10,tomar:bj2013,tomar:jpcb14},
in explicating the unanticipated importance of long-range interactions
in the role of denaturants \cite{tomar:gdmjcp18}, in revealing the
critical role of solute-solvent attractive interactions in biomolecular
hydration \cite{tomar:jpcb16,asthagiri:gly15}, and most recently, in
revealing breakthrough insights into decades-old assumptions about
hydrophobic hydration \cite{tomar:jpcl20}.

The molecular quasi-chemical theory is rooted in the
potential distribution theorem (PDT) \cite{widom:jpc82}. However,
instead of viewing the PDT from the more conventional lens of a
\emph{test particle method}, we view the PDT as defining a local
partition function \cite{lrp:apc02,lrp:book,lrp:cpms,lrp:ES99}. This
shift in perspective together with the concepts of conditional means and
the rule of averages defined below leads to the molecular
quasi-chemical theory of solutions. This approach naturally reveals
unexpected and clear connections to other theories of solution, an
aspect that we have noted in earlier reports
\cite{lrp:book,Zhang:jpcb14}, and is also seen in the work by other
groups \cite{Vafaei:jcp14}.  Importantly the molecular QCT
approach provides a rigorous and physically transparent framework to
\emph{conceptualize} and \emph{model interactions} in molecular solutions. It is this
latter perspective that we emphasize here. We present key ideas of the
theory and the results and refer the reader to the literature for
exhaustive details of the calculations. 

 \section{Theory}

The excess chemical potential, $\mu^{\rm (ex)}$,  is that part of the Gibbs free energy that arises from intermolecular interactions
and is the quantity of principle interest in understanding the solubility of a solute in water. All other properties of interest in understanding hydration are accessible once $\mu^{\rm (ex)}$ is known as a function of the solution conditions ($T, p,$ and composition). 

Formally, $\mu^{\rm (ex)}$ is given by the potential distribution theorem \cite{lrp:book,widom:jpc82} in one of two forms 
\begin{subequations}
\begin{eqnarray}
\beta\mu^{\rm (ex)} &  = & \ln \langle e^{\beta \varepsilon} \rangle \label{eq:pdta}\\ 
\textrm{or} \quad    \beta\mu^{\rm (ex)}    &   = & -\ln \langle e^{-\beta \varepsilon} \rangle_0 \label{eq:pdtb}
\end{eqnarray}
\label{eq:pdt}
\end{subequations}
where $\varepsilon = U_{N} - U_{N-1} - U_{\rm s}$ is the binding energy of the solute with the rest of the fluid. $U_{N}$ is the potential energy of the $N$-particle system at a particular configuration, $U_{N-1}$ is the potential energy of the configuration but with the
solute removed, and $U_{\rm s}$ is the potential energy of the solute.   In the first form, Eq.~\ref{eq:pdta}, the averaging $\langle\ldots\rangle$ is over $P(\varepsilon)$, the probability density distribution of $\varepsilon$ when both the solute and the solvent are thermally coupled; in the second form, Eq.~\ref{eq:pdtb}, the averaging $\langle\ldots\rangle_0$ is over $P^{(0)}(\varepsilon)$, the probability density distribution of $\varepsilon$ when the solute and the solvent are thermally uncoupled. (In Eqs.~\ref{eq:pdt}
and in the relations to follow, for simplicity we assume a solute in a fixed conformation. All these relations
are readily extended to include averaging over solute conformational states; for example, $\langle\ldots\rangle \rightarrow \langle\langle \ldots \rangle \rangle$. Here one set of brackets indicates averaging over solvent states and the other over solute conformational states.)

To appreciate the difficulties in using Eqs.~\ref{eq:pdt} directly, first consider Eq.~\ref{eq:pdta}. The exponential factor weights the higher energy configurations, but these are also the configurations that are poorly sampled when the solute and solvent are thermally coupled. Indeed, when the solute and solvent are coupled, the lower binding energy configurations are the ones that are better sampled. Likewise, to succeed in using Eq.~\ref{eq:pdtb}, better sampling of the lower energy configurations is required, but in practice only the high energy configurations are more accessible. It is precisely to alleviate these difficulties that one resorts to some form of alchemical 
approach wherein the solute-solvent interaction is scaled and the system progressively driven from the fully uncoupled state to the fully coupled state. While these alchemical methods can be quite robust \cite{chipotpohorille}, they necessarily involve unphysical solute states and can occlude the physical insights that are
of first interest. Further, on a methodological level, such unphysical solute states can bedevil all-electron calculations.

\subsection{Regularization}

Let $\chi$ be a mechanical variable, a function of the coordinates of the system. Then the average of $\chi$ in the coupled solute-solvent system 
is
\begin{eqnarray}
\langle \chi \rangle = \frac{ \langle \chi e^{-\beta \varepsilon}\rangle_0 }{\langle e^{-\beta \varepsilon} \rangle_0} \, .
\label{eq:average1}
\end{eqnarray}
This important result is the rule of averages.  (As a check, note that $\chi = \exp(\beta \varepsilon)$ immediately leads to Eq.~\ref{eq:pdta}, a highly nontrivial result.) Defining $\chi$ on the basis of physically meaning order parameters \cite{safirphd} is an important step in the overall development.

We now introduce the idea of conditional averages. For two propositions $A$ and $B$, Bayes' theorem teaches us that the probability of the joint $P(A\cdot B) = P(A) P(B|A)$. The same applies to the average of $\chi e^{-\beta \varepsilon}$. That is, 
the average of the joint is the average of $\chi$ times the average of $e^{-\beta \varepsilon}$ given $\chi$, as such 
\begin{eqnarray}
\langle \chi \rangle = \frac{ \langle \chi \rangle_0 \cdot \langle e^{-\beta \varepsilon} | \chi \rangle_{0} }  {\langle e^{-\beta \varepsilon} \rangle_0} \, , 
\label{eq:average2}
\end{eqnarray}
where the notation $\langle \ldots | \chi \rangle_{0}$ denotes averaging in an ensemble where the solute is thermally uncoupled from the solvent and the constraint is active. Eqs.~\ref{eq:average1} and~\ref{eq:average2} are what make the PDT \emph{a practical and economical} approach to the molecular theory of solutions \cite{lrp:book,lrp:cpms}.  

One particularly useful order parameter for hydration thermodynamics is the solute hydration state defined by the number $n$ of water molecules around the solute. Defining ``around" requires spatially partitioning the system into a local domain around the solute, the inner shell, and the domain outside, the outer shell.  We effect the partitioning by considering the constraint $\chi$ to be an indicator function which is unity when the inner shell is devoid of solvent and zero otherwise. Then $x_0 = \langle \chi \rangle$, the probability of observing an empty inner shell in the coupled solute-solvent system; $p_0 = \langle \chi \rangle_0$ is the probability of observing a cavity of the same extent as the inner shell in the neat solvent system; and $ \beta\mu^{\rm (ex)}(n=0) = -\ln \langle e^{-\beta \varepsilon} | \chi\rangle_{0}$ is the excess chemical potential of a solute that is conditioned to have an empty $n=0$ inner shell.

\begin{figure*}
\centering
\includegraphics[width=5.25in]{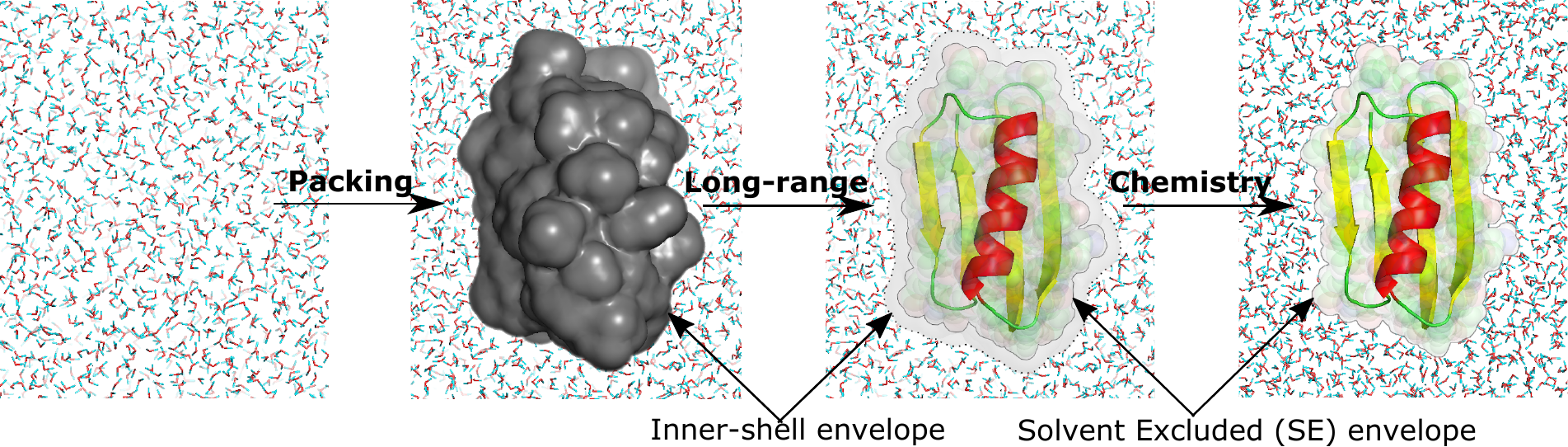}
\caption{Quasi-chemical organization of the excess chemical potential.  The inner-shell  identifies the region enclosing the solute for which the
solute-solvent binding energy distribution $P(\varepsilon|\phi)$ is accurately Gaussian. It approximately corresponds to the 
traditional first hydration shell of
the solute. The chemistry contribution is zero for the solvent-excluded envelope. The free energy to create the cavity to accommodate the solute gives the packing (hydrophobic) contribution. The chemistry and long-range parts determine the hydrophilic contributions. Reprinted from Ref.~\citenum{tomar:jpcl20}, copyright (2020) American Chemical Society.}\label{fg:cycle}
\end{figure*}

Rearranging Eq.~\ref{eq:average2}, we thus find 
\begin{eqnarray}
\beta\mu^{\rm (ex)} = \underbrace{- \ln p_0}_{\rm packing} + \underbrace{\beta \mu^{\rm (ex)}(n=0)}_{\rm long\, range} + \underbrace{\ln x_0}_{\rm chemistry} \, 
\label{eq:qc}
\end{eqnarray}
where $-\ln p_0$ is the free energy to create an empty cavity of the same extent as the inner shell in the solvent. This packing contribution is precisely the primitive hydrophobic contribution to hydration. $\ln x_0$ is the free energy gained in allowing solvent to populate an empty shell. This quantity  assesses the contributions of short-range, specific interactions that can be characterized as chemical in nature; we call this the chemistry contribution. The excess chemical potential of the conditioned solute gives the long-range contribution to hydration.  The chemistry and long-range contribution together give the hydrophilic contribution to hydration. Eq.~\ref{eq:qc} has a simple schematic 
interpretation as shown in Fig.~\ref{fg:cycle}. 

To see what we have achieved in going from Eqs.~\ref{eq:pdt} to Eq.~\ref{eq:qc}, by introducing an external constraint $\chi$, the statistically ill-behaved problem (Eqs.~\ref{eq:pdt}) has been recast as a problem involving the calculation of three well-behaved contributions. In particular, in $\beta\mu^{\rm (ex)} \rightarrow \beta\mu^{\rm (ex)}(n=0)$, since
we move the solute-solvent boundary outward, the solute-solvent interaction is tempered. We can chose the inner-shell domain just large enough that the conditioned binding energy distribution $P(\varepsilon | n=0)$ is gaussian, simplifying the evaluation of $\beta\mu^{\rm (ex)}(n=0)$. That practical simplification has a deep physical significance: a gaussian binding energy distribution arises from small (relative to thermal energies) \emph{non-specific} contributions, and by construction, the corresponding $\ln x_0$ captures all the effects of \emph{specific} solute-solvent interactions. Thus \emph{theory teaches us} what interactions are chemically significant and what are not in considering the hydration of a solute. 

Summarizing the development so far, the constraint $\chi$ helps to \emph{regularize} the statistical problem posed by Eqs.~\ref{eq:pdt} and in that process also teaches us about the physically meaningful partitioning of the hydration thermodynamics of the solute. We do face the problem of evaluating $x_0$ and $p_0$. Before we consider direct numerical approaches to $x_0$ and $p_0$, we highlight the partition-function perspective of these quantities, a perspective that has played an important role in studies on hydration of solutes ranging from ions to small, prototypical hydrophobes. 

\subsection{Quasi-chemical organization}\label{sec:QC}
Consider the chemical reaction, 
\begin{eqnarray}
\textrm{X} + n \textrm{H$_2$O} \rightleftharpoons \textrm{X[H$_2$O]}_n
\end{eqnarray}
involving the binding of $n$ water molecules to the empty inner-shell of the solute X. Let $x_n$ be the fraction of solute with $n$ water molecules in the inner shell; $\rho_w$ the bulk density of water; and $K_n$ the equilibrium constant for the reaction. We then have 
$x_n / x_0 = K_n \rho_w^n$, and thus
\begin{eqnarray}
\ln x_0 = -\ln \left[1 + \sum\limits_{n \geq 1} K_n \rho^n \right]
\label{eq:x0expansion}
\end{eqnarray}
One can construct a similar expansion for $p_0$. It is this chemical organization of $x_0$ and $p_0$ that leads to the appellation `quasi-chemical' applied to Eq.~\ref{eq:qc}. The equilibrium constants are themselves configurational integrals, i.e.\ partition functions, but for the few-body problem of $n$ solvent molecules in the inner-shell. Reducing the complexity to few-body problems allows one to use high-level quantum chemical methods with full account of multi-body interactions. Of course, care is needed in accounting for the role of the environment outside the inner 
shell \cite{lrp:hspre,merchant:jcp11b}, a topic we shall address in the section on future directions below.

\subsection{Multi-state organization} 
In deriving Eq.~\ref{eq:qc}, we assumed that the inner-shell was devoid of solvent. However, the development holds for any 
coordination state $n$. Thus, for example, the constraint $\chi$ can be an indicator function that selects only cases where $n$ solvent molecules are in the inner shell. This then leads to \cite{merchant:jcp09,asthagiri:cpl10}
\begin{eqnarray}
x_n = p_n \cdot e^{-\beta[\mu^{\rm (ex)}(n) - \mu^{\rm (ex)}]} \, ,
\label{eq:qcgeneral}
\end{eqnarray} 
a relation that transparently shows how solute-solvent interactions codified by the exponential term modulate the intrinsic properties of the solvent ($\{p_n\}$) to give rise to the distribution of coordination states around the solute ($\{x_n\}$). The role intrinsic properties of the solvent plays in determining the coordination state of the solute is an aspect that is not often well-appreciated, an aspect that we consider in the applications below. 

Since the occupation probabilities $\{x_n\}$ and $\{p_n\}$ are normalized, we immediately find that 
\begin{subequations}
\begin{eqnarray}
\beta \mu^{\rm (ex)} & = & \ln \sum_{n} x_n e^{\beta \mu^{\rm (ex)}(n)}  \\
\beta \mu^{\rm (ex)} & = & -\ln \sum_{n} p_n e^{-\beta \mu^{\rm (ex)}(n)}
\end{eqnarray}
\end{subequations}
which are the multi-state generalizations of Eqs.~\ref{eq:pdta} and~\ref{eq:pdtb}, respectively.  These relations provide insights
on how individual coordination states contribute to the net excess chemical potential of the solute.

\subsection{Direct evaluation of $x_0$ and $p_0$}

For molecular solutes and for macromolecules,  obtaining $x_0$ (or $p_0$) by calculating the $n$-body equilibrium constant appearing in Eq.~\ref{eq:x0expansion} is a formidable challenge. One faces similar challenges in interpreting \emph{ab initio} molecular
dynamics simulations. For these cases, we facilitate  the calculation of $x_0$ (or $p_0$)
by introducing an auxiliary field $\phi(r;\lambda)$ that moves the solvent away from the solute, thereby tempering the solute-solvent binding energy.   Thus instead of the hard constraint,  the inner-outer boundary is defined by a soft-boundary \cite{lrp:softcutoff,weber:jcp11}. (If required, the soft-cavity results are easily corrected to give the hard-cavity result \cite{weber:jcp11}.) As above, the conditional distribution $P(\varepsilon|\phi)$ is better characterized than $P(\varepsilon)$, and more importantly, in calculations we can adjust $\lambda$, the range of
the field, to control approximation of $P(\varepsilon|\phi)$ as a 
Gaussian distribution.

With the introduction of that auxiliary field \cite{Weber:jctc12,tomar:bj2013,tomar:jpcb14,tomar:jpcb16,asthagiri:gly15,tomar:gdmjcp18,asthagiri:jpcb20a,tomar:jpcl20}
\begin{eqnarray}
\beta \mu^{\mathrm{(ex)}} = 
\underbrace{- \ln p_0[\phi]}_{\rm packing} + 
\underbrace{\beta\mu^{\mathrm{(ex)}} [P(\varepsilon|\phi)]}_{\rm long-range} + 
\underbrace{\ln x_0[\phi]}_{\rm chemistry}~,
\label{eq:qcPhi}
\end{eqnarray}
which is Eq.~\ref{eq:qc} but where the constraint $\chi$ is now an auxiliary field.
Thus the individual contributions are functionals of the auxiliary field, as indicated. 

For the field defined by the repulsive part of the WCA potential for water-water interactions, for the hydration of biomolecules \cite{tomar:bj2013,tomar:jpcb14,tomar:jpcb16,asthagiri:gly15,tomar:gdmjcp18,asthagiri:jpcb20a,tomar:jpcl20}
we find that $\lambda \approx 5${\AA}  ensures that the conditional binding energy distribution is accurately Gaussian, and we denote this range as $\lambda_\mathrm{G}$.  The largest value 
of $\lambda$ for which the chemistry contribution is
negligible,
labeled $\lambda_{\rm SE}$, has a special meaning. It bounds the domain excluded
to the solvent.  We find
$\lambda_{\rm SE} \approx 3${\AA}, unambiguously; see also Refs.~\citenum{tomar:jpcb16,asthagiri:gly15,asthagiri:jpcb20a}.  With this 
choice, Eq.~\eqref{eq:qc} can be rearranged as, 
\begin{eqnarray}
\beta \mu^{\mathrm{(ex)}}  =   
\underbrace{-\ln p_0(\lambda_{\rm SE})}_{\rm solvent\, exclusion} +   
\underbrace{\beta\mu^{\mathrm{(ex)}}[P(\varepsilon|\lambda_\mathrm{G})]}_{\rm long-range} + 
\underbrace{ \ln \left[x_0(\lambda_\mathrm{G})\left(\frac{p_0(\lambda_{\rm SE})}{p_0(\lambda_\mathrm{G})}\right)\right]}_{\rm revised\, chemistry}~.
\label{eq:qc1}
\end{eqnarray}

The several contributions are
identified by the range parameter. Thus, for example,
$x_0(\lambda_\mathrm{G})\equiv x_0[\phi(\lambda_\mathrm{G})]$. 
The revised chemistry term has the physical
meaning of the work done to move the solvent interface a distance
$\lambda_\mathrm{G}$ away from the 
volume excluded by the solute relative to the case when the
only role played by the solute is to exclude solvent up to $\lambda_{\rm
SE}$.  This term highlights the role of \emph{short-range} solute-solvent
attractive interactions on hydration. Interestingly, the range between
$\lambda_\mathrm{SE} = 3${\AA} and $\lambda_\mathrm{G} = 5${\AA}
corresponds to the first hydration shell for a methane
carbon \cite{asthagiri:jcp2008} and is an approximate descriptor of the
first hydration shell of groups containing nitrogen and oxygen heavy
atoms.

\subsection{Enthalpy and entropy of hydration}
Once we determine the excess chemical potential, we can obtain both the hydration entropy and the hydration enthalpy. 
The excess entropy of hydration is  given by \cite{tomar:jpcl20}
\begin{eqnarray}
T  s^{\mathrm{(ex)}}  = 
E^{\mathrm{(ex)}}  + p\left\lbrack\left\langle V^{\mathrm{(ex)}} \right\rangle + kT \kappa_T\right\rbrack - \mu^{\mathrm{(ex)}} - kT^2 \alpha_p 
\label{eq:entropy}	              
\end{eqnarray}
where $\kappa_T$ is the isothermal compressibility of the solvent,
$\alpha_p$ is the thermal expansivity of the solvent,  and $\langle
V^{\mathrm{(ex)}} \rangle$ is the average excess volume of 
hydration.   (Of course, the excess entropy can also be obtained from a temperature derivative of $\mu^{\rm (ex)}$, thereby providing 
a rigorous consistency check of the calculations \cite{tomar:jpcl20}.) For discussion below, we write
$E^{\mathrm{(ex)}} = E_{\mathrm{sw}} + E_{\mathrm{reorg}}$
with $E_{\mathrm{sw}}$ the average solute-water
interaction energy  and $E_{\mathrm{reorg}}$  
the reorganization energy.  Ignoring small pressure-volume effects
leads to enthalpies
\emph{e.g.,} $h^{\mathrm{(ex)}} =  E^{\mathrm{(ex)}}$, 
and the broader outlook
\begin{eqnarray}
T  s^{\mathrm{(ex)}} \approx  h_{\mathrm{sw}} + h_{\mathrm{reorg}}  - \mu^{\mathrm{(ex)}}~.
\label{eq:entropy2}	              
\end{eqnarray}
The temperature derivative of the enthalpy (or the entropy) immediately gives the heat capacity. 

The above development has opened new avenues to interrogate the hydration thermodynamics of macromolecules at the same
level of resolution as that possible for a small molecule, such as methane. The impact of this transformative development
will be discussed in the applications below.

\section{Applications}
The quasi-chemical approach sketched above and its further specializations have been used to probe the structure and thermodynamics 
underlying the hydration of monoatomic solutes \cite{Martin:jpca98,rempe:lijacs,rempe:nafpe,grabowski:jpca02,Ashbaugh:kr03,Asthagiri:becpl,Asthagiri:hocpl,asthagiri:jcp03,Asthagiri:pnas04,Asthagiri:pnas05,rempe:kpccp,rempe:bc06,beck:jcp08,asthagiri:jcp2008,merchant:jcp09,beck:jcp10,beck:jpcb11,asthagiri:cpl10,merchant:jcp11b}, water \cite{Asthagiri:pre03,paliwal:jcp06,shah:jcp07,lrp:jpcb09,weber:jcp10a,weber:jcp10b,weber:jcp11,merchant:jcp11a}, small multi-atomic solutes \cite{asthagiri2007non,paulaitis:corr10,asthagiri:jpcb20a}, short peptides \cite{tomar:bj2013,tomar:jpcb14}, and longer peptides and macromolecules \cite{Weber:jctc12,asthagiri:gly15,tomar:gdmjcp18,asthagiri:jpcb20a,tomar:jpcl20}.  These solutes span a diversity of chemical complexity ranging from those that chemically interact with water \cite{Martin:jpca98,rempe:lijacs,grabowski:jpca02,Asthagiri:becpl,Asthagiri:hocpl,asthagiri:jcp03,Asthagiri:pnas04,Asthagiri:pnas05}, to cases
where specific chemical bonding may be ignored but one still needs to consider strong short-range interactions \cite{rempe:nafpe,rempe:kpccp,rempe:bc06,shah:jcp07,beck:jcp08,shah:jcp07,merchant:jcp09,beck:jcp10,beck:jpcb11,asthagiri:cpl10,merchant:jcp11a,merchant:jcp11b}, to prototypical hydrophobes \cite{Ashbaugh:kr03,asthagiri:jcp2008,asthagiri:jpcb20a}, and 
to solutes which are heterogenous in terms of their chemical composition\cite{paulaitis:corr10,Weber:jctc12,tomar:bj2013,tomar:jpcb14,asthagiri:gly15,tomar:gdmjcp18,asthagiri:jpcb20a,tomar:jpcl20,paulaitis:corr21}. The theory has been used to enhance the interpretation of \emph{ab initio} simulations \cite{rempe:lijacs,rempe:nafpe,Asthagiri:pre03,Asthagiri:pnas04,Asthagiri:pnas05,rempe:kpccp,weber:jcp10a,weber:jcp11} which are 
necessarily limited to short time scales, small system sizes, and the inherent limitation of the underlying electron density functionals.  The theory has also been used to study non-aqueous systems such as hard-sphere fluids \cite{lrp:jpcb01,lrp:hspre}, mixed solvents \cite{Tam:2012jo}, and in suggesting ways to incorporate multi-body effects in the description of associating patchy-colloids \cite{Bansal:2017iv}. These developments 
have been pursued primarily by groups led by the authors here, by Rempe and her collaborators at Sandia National Laboratory,
and by Beck and his collaborators at the University of Cincinnati.  There has also been an attempt using a different way of sampling hydration states  \cite{rxp} to model the hydration free energy of a  macromolecule within a less physically transparent, QCT-like framework.

The above outline already shows a great diversity of applications, but it is still rather incomplete. For brevity, we consider four examples  below.  In these examples, QCT in concert with simulations leads to \emph{singularly incisive} 
insights that help deepen our understanding and appreciation of problems of long-standing interest in aqueous phase chemistry and biology. 


\subsection{Hydration of K$^+$}\label{sc:K+}

Understanding the hydration structure of ions is a basic requirement in developing a molecular scale theory of aqueous electrolytes, and, in the
 case of K$^+$, learning how it interacts with biological material such as ion channels. In the context of the potassium channel it had been proposed that the channel selects for K$^+$ over Na$^+$ because the eight (8) carbonyls in the selectivity filter of the channel better 
compensate for the dehydration of K$^+$ over Na$^+$.  Comparing the population distribution of water molecules in the first hydration shell of the ion would indeed suggest that the eight (8) coordinate structure of K$^+$ is more probable than the eight (8) coordinate structure of Na$^+$, in 
apparent accord with the proposed model of selectivity.  Investigations founded on the quasi-chemical approach however uncovers subtleties that ultimately leads one to reject the proposed model of selectivity.  We consider
this example below. 

\begin{figure}
\includegraphics[width=3.125in]{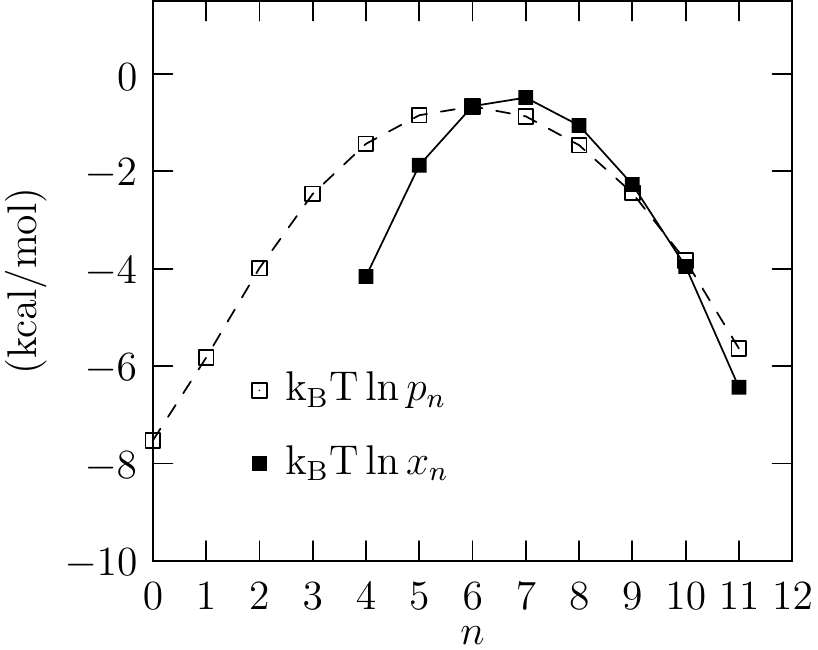}\\
\includegraphics[width=3.125in]{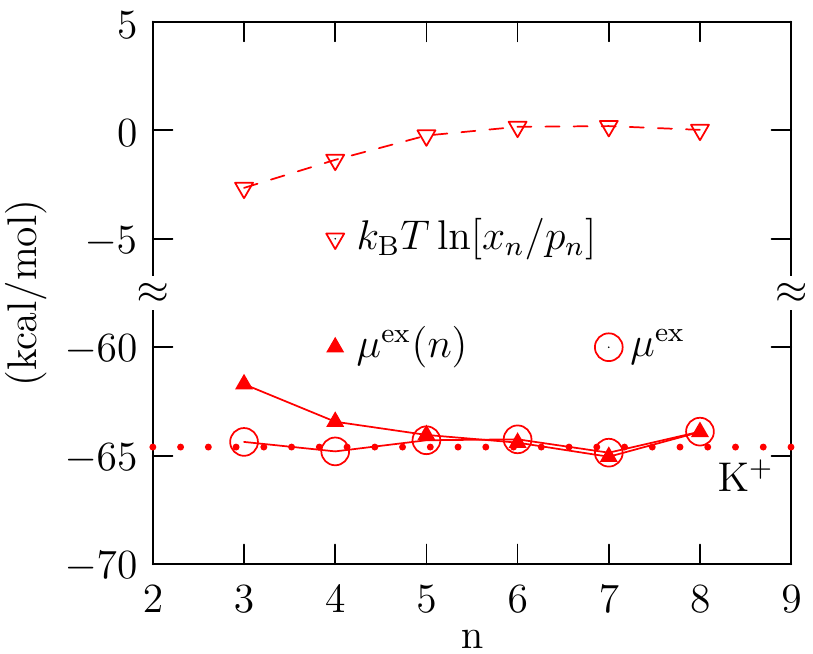}
\caption{Top panel: Coordination number distribution in the presence $\{x_n\}$ and absence $\{p_n\}$ of the ion in a coordination volume of radius 3.5~{\AA}. Bottom panel: Comparison of the excess chemical potential of the ion in a given coordination state with the net excess chemical potential obtained on the basis of the quasi-chemical theory. The dotted line is the independent coupling parameter estimate of the hydration free energy of the ion. Note that by $n=5$, we recover the net hydration free energy of the ion. Adapted from Figures 2 and 3 of Ref.~\citenum{asthagiri:cpl10}, with permission from Elsevier (copyright 2010).}\label{fg:Kxnpn}
\end{figure}

Figure~\ref{fg:Kxnpn} (top panel) shows the probability distribution of waters of hydration in a coordination shell of radius 3.5~{\AA} around K$^+$.
These simulations show that the mean coordination number of K$^+$ is 6.8 water molecules; the most probable coordination number is 7; and the probability of the $n=8$ state is about 1/3$^{\rm rd}$ the probability of the $n=7$ state. Thus, the analysis of coordination around the ion would would lend support to the hypothesis noted above. 

Theory teaches us that the occupation distribution of the observation volume in the absence of the ion, $\{p_n\}$, is a crucial ingredient in understanding the coordination distribution in the presence of the ion, $\{x_n\}$ (Eq.~\ref{eq:qcgeneral}). Relative to the $\{p_n\}$ distribution, the behavior of the coordination state of the ion above and below the most probable coordination state, $n=7$, are strikingly different. For $n \geq 6$, the $\{x_n\}$ distribution closely tracks the $\{p_n\}$ distribution, whereas the deviations are pronounced for $n < 6$. From Eq.~\ref{eq:qcgeneral} we then find that 
for coordination states $n \geq 6$, the excess chemical potential of the ion is insensitive to the coordination; said differently, coordination states up to $n=6$ are enough to determine the hydration free energy of the ion. 

This example illustrates the insights QCT uncovers. First, the dominant coordination states, i.e.\ those states that determine the net free energy, typically lie \emph{below} the most probable coordination state. Second, the most probable coordination state already begins to reflect the intrinsic density fluctuations of water at the scale of the coordination volume. As a corollary of these two observations, only a small number of water molecules actually sense the chemical type of the ion, an insight that is useful in understanding specific-ion effects in aqueous phase processes.

\subsection{Hydration of the peptide backbone}\label{sc:bb}

The extant understanding of protein solution thermodynamics has been strongly influenced by studies on model compounds. In this spirit, to understand the role of the solvent on the peptide backbone, researchers earlier studied short segments of (Gly)$_n$ residues, where $n$ is the chain length. It both experiments and simulations the free energy of hydration is found to be linear in $n$; experimentalists use this finding to impute the slope --- the incremental change in free energy per additional residue --- as the contribution of the peptide backbone. Investigations founded on the quasi-chemical approach reveals the flaw in this reasoning, one with significant consequences for how we should think about protein solution thermodynamics. 

\begin{figure}
\includegraphics[width=3.125in]{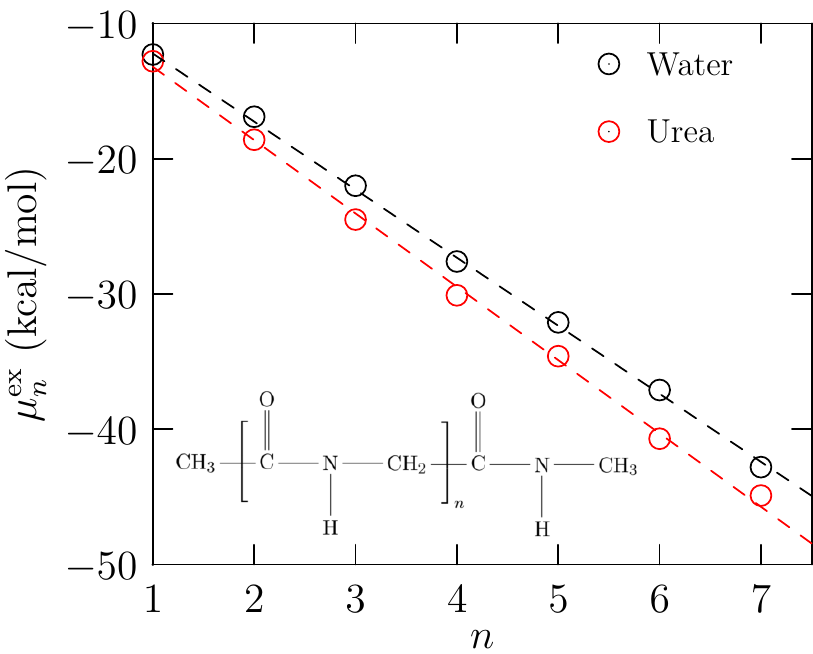}
\caption{Hydration free energy of (Gly)$_n$ in an extended conformation for various degree of polymerization. Adapted from Fig.~3 of Ref.~\citenum{tomar:bj2013} with permission from Elsevier (copyright 2013).}\label{fg:glyn}
\end{figure}

Figure~\ref{fg:glyn} shows the hydration free energy of (Gly)$_n$ in an extended conformation in both water and in a 8 M aqueous urea solution. 
At first glance the striking linearity seemingly supports the idea of attributing a constant free energy contribution per additional peptide group.  The difference in the slopes in aqueous urea and water is then seen as the contribution of urea to the differential solvation of the peptide group. The same behavior is seen for the individual contributions in Eq.~\ref{eq:qcPhi}. But does this imply that free energy contributions due to individual peptide units are additive? 

Drawing upon the QCT theory, we partition the peptide into the $i^{\rm th}$ group and the rest ($i_{back}$ for background). Then $p_0$ for the entire peptide should be $p_0 = p_0(i) p_0 (i_{back}) p_c$, where $p_c$ is the correlation. If additivity holds, then $\log p_c = 0$. (Similar comments apply to $x_0$ and the long-range contribution.) Table~\ref{tb:xpcorr} shows that the correlation contributions are not zero but negative, indicating cooperativity. This cooperativity of hydration can also be seen by analysing binding energy distributions \cite{paulaitis:corr10,tomar:jpcb14,paulaitis:corr21}. The correlation contributions are relatively constant, and it is this feature that masks the breakdown of additivity!   

\begin{table}[h!]
\caption{Correlation contributions to the net chemical plus packing contribution. Always $i$ is the central peptide unit; $n$ is the degree of polymerization. The $x_0/p_0$, $x_0(i)/p_0(i)$, and $x_0(i_{back})/p_0(i_{back})$ values are reported in energy units. The standard error on $k_{\rm B}T \ln x_c / p_c$ is about 0.6~kcal/mol. Within this uncertainty, the correlation contribution is the same for both urea and water. All values are in kcal/mol. Reprinted from Table 5 of Ref.~\citenum{tomar:bj2013} with permission from Elsevier (copyright 2013).}\label{tb:xpcorr}
\centering
\begin{tabular}{llcccc} 
solvent &  $n$ & $\displaystyle  \frac{x_0}{p_0}$ &  $\displaystyle \frac{x_0(i)}{p_0(i)}$ &  $\displaystyle\frac{x_0(i_{back})}{p_0(i_{back})}$ & $\displaystyle\frac{x_c}{p_c}$ \\ \hline
Water &    4    &  $-12.4$ & $3.3$ & $-7.4$ & $-8.3$ \\ 
            &  6     &  $-17.2$ & $3.2$ & $-12.9$ & $-7.5$ \\ \hline
Urea   &     4   &  $-13.0$    & $3.4$  & $-7.9$ & $-8.5$ \\ 
            &  6    &  $-18.5$ & $3.5$ & $-14.5$ & $-7.5$ \\ \hline   
\end{tabular}
\end{table}

\subsection{Temperature dependence of hydration free energies}\label{sc:tempeffects}

Reversible unfolding of a soluble protein molecule upon cooling is a typical example of \emph{cold denaturation}. The molecular structure then refolds upon heating, and the  higher temperature system --- with the  functionally \emph{folded} protein molecule --- has the higher entropy. It is striking that this so-called \emph{inverse} temperature behavior is exhibited in the hydration of a non-polar solute such as CH$_4$. Thus, with some notable exceptions, cold denaturation suggests another confirmation of the significance  of hydrophobic effects in protein stability. But is this so? 

The molecular mechanism of this inverse temperature behavior  has been the subject of 
debate since the earliest days of statistical mechanical theory of hydrophobic 
effects \cite{pratt1980effects,rossky1980benzene,pratt1985theory,asthagiri:jcp2008}.  The
most basic question was whether this inverse temperature behavior derives
essentially from packing, i.e.\ van~der~Waals excluded volume interactions, and the consequent 
structures.  Here again theory proves transformative. Using QCT we can determine whether the packing contribution dominates, but more importantly, we can straightforwardly compare that contribution to contributions from hydrophilic interactions to unambiguously resolve this long-standing question. Inverse temperature behavior is typically expressed in all 
the contributions to hydrophobic free energies, but typically
preeminently in long-ranged contributions 
(Fig.~\ref{fg:cycle}) \cite{chaudhari:jpcb15,pratt2016statistical,gao2018role,tomar:jpcl20}. 

\begin{figure*}
\includegraphics[width=6in]{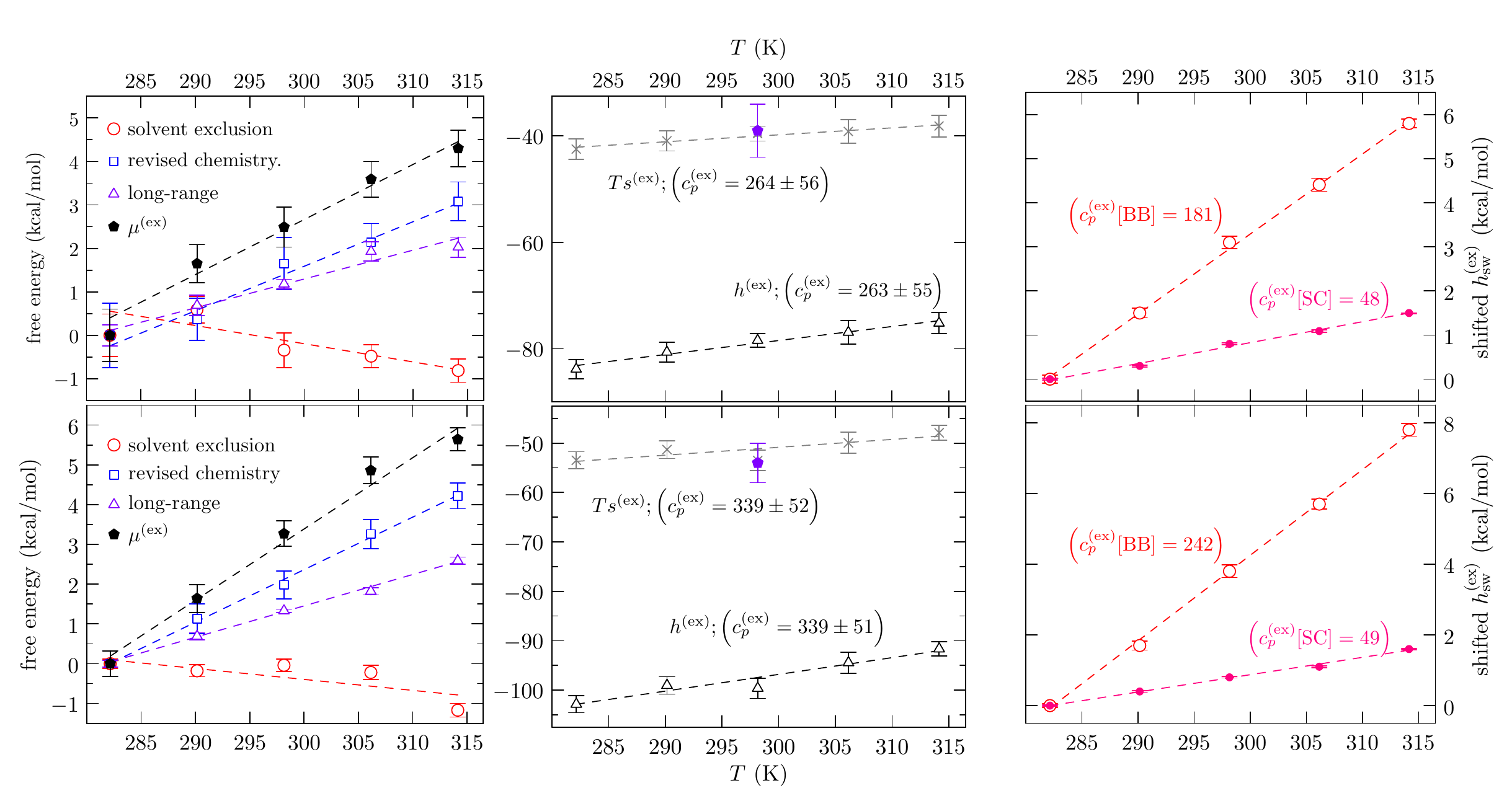}
\caption{Hydration thermodynamics of a deca-alanine helix (top row) and corresponding coil (bottom row). All the values are noted 
 \underline{\emph{relative}} to the $T = 282.15$~K value. The rightmost panel shows the separate contributions 
to solute-water enthalpy $h^{\mathrm{(ex)}}_{\rm sw} = h^{\mathrm{(ex)}}_{\rm BB} +
h^{\mathrm{(ex)}}_{\rm SC}$.  $h^{\mathrm{(ex)}}_{\rm BB}$ is the
contribution from backbone-solvent interactions and
$h^{\mathrm{(ex)}}_{\rm SC}$ is the contribution from side chain-solvent interactions. $Ts^{\rm (ex)}$ computed as 
as the temperature derivative of $\mu^{\rm (ex)}$ (middle panel, purple symbol) is in excellent agreement with the value obtained using Eq.~\ref{eq:entropy2}; likewise the heat capacities obtained using either $s^{\rm (ex)}$ or $h^{\rm (ex)}$ agree, providing a strong consistency check of the calculations. Reprinted from Fig.~3 of Ref.~\citenum{tomar:jpcl20}, Copyright(2020) American Chemical Society.}\label{fg:cold}
\end{figure*}

Studies on polypeptide hydration founded on the QCT theory had already shown the importance, indeed dominance, of hydrophilic effects in protein hydration \cite{tomar:jpcb16}. That line of research culminated in a first-of-its-kind analysis (Fig.~\ref{fg:cold}) that reveals surprisingly new insights that challenges dogma built over the decades \cite{tomar:jpcl20}. 

First, consider the packing (or hydrophobic contribution). Notice that theory shows that the packing contribution for the polypeptide in fact decreases with increasing temperature. Thus hydrophobic hydration \emph{does not become stronger} with increasing temperature. This is in contrast to what is seen for methane, where hydrophobic contributions become stronger with increasing temperature, a striking example of the breakdown of additivity.  

Second, consider the chemistry together with the long-range contributions. These hydrophilic contributions become weaker with increasing temperature. Moreover, these hydrophilic contributions \emph{dominate} the hydrophobic contributions, such that the temperature dependence of the net chemical potential follows the trend set by the hydrophilic contributions. That temperature dependence would suggest that the entropy of hydration is negative, but as the results plainly show, this negative negative entropy of hydration and the positive heat capacity of hydration both arise from \emph{hydrophilic effects}. Further, as the results in Ref.~\citenum{tomar:jpcl20} show, there is no need to invoke specific structural arrangements of water to rationalize the negative entropy and positive heat capacity of hydration.  Thus, the insights emerging from the QCT approach encourage \emph{a paradigm shift} from dominant effects to compensating effects in bio-macromolecular folding and assembly.

\subsection{System size effects in simulations}\label{sc:systemsize}

In the final example, we discuss how QCT has helped uncover technical subtleties in simulating macromolecules. The problem was motivated by our inconclusive efforts understanding the conformational switch in protein $G_B$. This protein
is remarkable in switching conformations from a $4\beta + \alpha$ fold to a $3\alpha$ fold upon mutation of a single residue \cite{bryan:pnas09}. In pursuing the reasons for the initial inconclusive result QCT revealed a rather important role played by system sizes in bio-macromolecular simulation, understanding of which would be essential in both designing and interpreting simulations.

The set $\{p_n\}$ of occupancy distribution in a cavity of size $v_o$ in a liquid can be described by its moments.  The first moment gives the mean occupancy $\bar{n} = \rho v_o$, where $\rho = N/V$ is the density of the fluid, $N$ is the total number of particles in the system and $V$ the (average) volume. The first moment will not depend on the system size provided the density is constant.  The second (central) moment gives the mean-squared fluctuation, $\delta^2$.  For a gaussian distribution of $\{p_n\}$, a key discovery in understanding hydrophobicity \cite{garde:prl96,lrp:jpcb98} was 
\begin{eqnarray}
-k_{\rm B}T \ln p_0 = k_{\rm B}T \frac{\rho^2 v_o^2}{2\delta^2} + \frac{k_{\rm B}T}{2} \log 2\pi\delta^2 \, .
\label{eq:dGit}
\end{eqnarray}

$\delta^2(n,v_o; N)$  is related to the pair distribution function $g({\bf{r}};N)$ in the $N$-particle system and depends on both $v_o$ and $N$.  Drawing upon the work of Lebowitz and Percus \cite{lebowitz:61a,lebowitz:61b}, Rom\'an et al.\ \cite{velasco:97} derived an expression for $\delta^2(n,v_o; N)/\bar{n}$. When the box volume $V >> v_p$, the volume of a solvent particle, and neglecting terms containing $(v_p/V)^{1/2}$ and $v_p/V$, we have from their work 
\begin{eqnarray}
\frac{ \bar{n} }{\delta^2 } \approx (1 + \frac{v_o}{V}) f(\eta) \, ,
\label{eq:nbar}
\end{eqnarray} 
where $f(\eta) = 1 + 4\eta + O(\eta^3)$ and $\eta = Nv_p/V$ is the packing fraction. Using Eq.~\ref{eq:nbar} in Eq.~\ref{eq:dGit} and ignoring the $\log \delta^2$ factor gives 
\begin{eqnarray}
-k_{\rm B}T \ln p_0  \approx \frac{ \bar{n} k_{\rm B}T }{2} (1 +  \frac{v_o}{V}) f(\eta) \, .
\label{eq:ss}
\end{eqnarray}
This relation shows that as $V$  increases the system can accommodate fluctuations better and the work to create a cavity in the liquid decreases. Thus the assessed magnitude of hydrophobic hydration will be smaller in a larger system. From Eq.~\ref{eq:qc}, assuming that the long-range interactions are only weakly dependent on $V$, we can infer that  $k_{\rm B}T \ln x_0 \propto -{v_o}/{V}$. Thus the assessed magnitude of (favorable) short-range hydrophillic contributions to hydration will be smaller (i.e.\ weaker) in a larger system. 

\begin{figure*}
\includegraphics[width=6in]{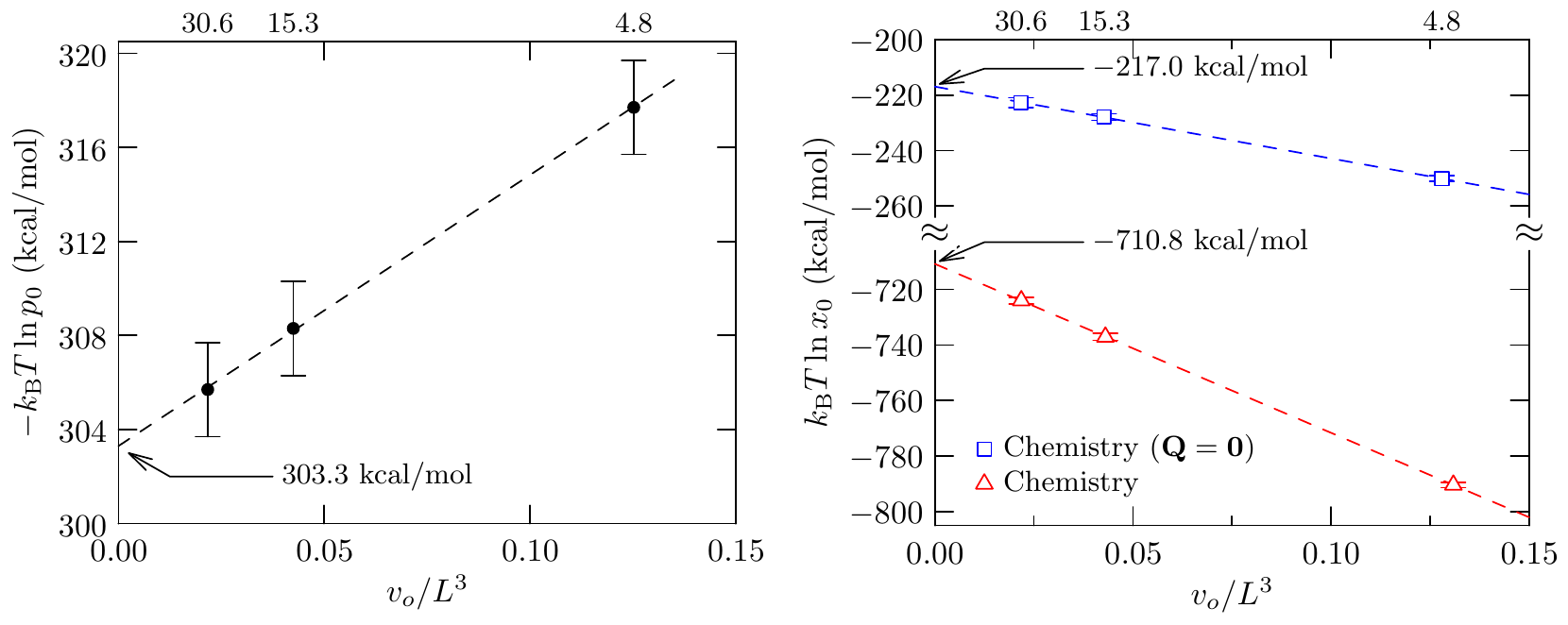}
\caption{The free energy to evacuate the hydration shell around $G_B$ (right panel) and to create the corresponding cavity in the liquid (left panel)
as a function of the size of the simulation cell. The top axis shows the number of water molecules (in thousands). Energies are in kcal/mol. Reprinted from Fig.~8 of Ref.~\citenum{asthagiri:jpcb20a}, Copyright(2020) American Chemical Society.}\label{fg:gb}
\end{figure*}

Figure~\ref{fg:gb} shows the $\ln p_0$ and $\ln x_0$ contributions in the hydration of $G_B$ and its analog with all partial charges turned off ($\textbf{Q}=0$). 
Just as theory predicts, the positive hydrophobic contribution to hydration becomes weaker or \emph{less stabilizing} of the folded conformation as the simulation system size is increased in size. 
(As an aside, this finding is in contrast to the suggestion that hydrophobic effects are strengthened in a larger simulation cell \cite{Meuwly:2018}.)
The negative hydrophilic contribution also becomes weaker or \emph{more stabilizing} of the folded conformation as the system size is increased. The net chemical potential is also dependent on the system size, but because of the competing roles of packing and hydrophilic contributions, the dependence is necessarily weaker than that displayed by the individual components. This is similar in spirit to what is commonly noted as entropy-enthalpy compensation. Importantly, the net free energies are 
dominated by the behavior of hydrophilic and  \emph{not the hydrophobic contributions}, emphasizing again the inversion of the hydrophobic-hydrophilic paradigm.

\section{Conclusions and Future Directions}  

Molecular quasi-chemical theory of solutions is both a \emph{practical framework} to obtain free energies from molecular simulations and also a \emph{generator} of models of molecular solutions. In this Perspective, we have highlighted the former aspect, while the latter aspect has been discussed in-depth in a monograph \cite{lrp:book}. 

As a practical framework, the theory allows one to use extant simulation approaches to interpret and model the solution thermodynamics of species ranging in complexity from simple mono-atomic solutes to complex, chemically heterogeneous macro-molecules. The theory has in-built consistency checks, such as the idea of cluster-variation, an idea with roots in the theory of phase transitions. Thus, QCT demands, and calculations show that, the results of the calculation are insensitive to how one partitions the system into local and long-range contributions. The theory shows that it proves illuminating to partition the system such that the consequences of attractive solute-solvent interactions are parsed into chemically specific, short-ranged and non-specific, long-range contributions. The latter admits a gaussian model and aids in simplifying the calculations. This partitioning also automatically guides us in the choice of the molecular envelope that is physically relevant in discussions of packing (or hydrophobic) contributions.

When applied to macromolecules, the QCT studies reveal the breakdown of additivity, a lynchpin assumption in experimental and theoretical studies on the solution thermodynamics of macromolecules. Thus, for example, studying the hydration of peptide backbones of various lengths, while of intrinsic interest, is not useful in isolating the hydration free energy of a single backbone unit. More dramatically, QCT reveals that  the hydration of a single cavity and that of a collection of cavities in the shape of the peptide have different temperature dependencies. Thus, it is fundamentally incorrect to extrapolate from the hydrophobic hydration of simple solutes to interpret the hydrophobic hydration of a macromolecule. 

Applications of QCT described here lead to the finding of the importance, indeed the dominance, of hydrophilic effects in the hydration thermodynamics of polypeptides.  This finding also provides a parsimonious explanation 
of cold denaturation: as temperature is reduced, the enhanced hydrophilic interaction between solvent water and the protein serves to pry apart the protein. This same understanding concisely explains the negative entropy and positive heat capacity signatures that are often attributed to hydrophobic hydration, since these same signatures are found in the hydration of prototypical, small molecule hydrophobes.  These initial studies were founded on simple, yet realistic, polypeptides.  The next steps in this line of work will require one to study more complex proteins. 

QCT explicitly considers the variation in the distribution of solvent molecules in an empty cavity (in the size and shape of the inner hydration shell of a solute) and in the inner shell of a solute. It turns out that the variation in occupancy is sensitive to the size of the simulation cell. Importantly, the (unfavorable) hydrophobic effects (that tend to collapse a polypeptide chain) and (favorable) hydrophilic effects (that help unravel a folded protein) are enhanced in a smaller simulation cell.  Understanding this balance will prove crucial in better modeling macromolecules and macromolecular complexes. A further challenge that can yield to theory is in modeling proteins that can switch conformations. Here, the multistate generalization of QCT will likely prove helpful. 

In the hydration of highly charged ions, in the primitive QCT formulation one ignores the role of the bulk medium and treats the ion-water cluster using modern electronic structure methods. This approximation  has proven remarkably successful, but for treating molecular and/or soft ions it would be essential to account bulk medium effects. Early studies on hard-sphere systems suggest that the bulk medium could be described using molecular fields, but this idea needs to be better developed for molecular liquids such as water. 

The system-size dependence uncovered by QCT will also be important in studies that are based on \emph{ab initio} (AIMD) simulations. Such simulations are necessarily limited to small system sizes and short times. The ideas developed in this Perspective may prove helpful in more careful and sensitive analysis of coordination state distributions found in such AIMD simulations.   
The QCT calculation approaches highlighted in this Perspective have been adapted within well-established packages such as NAMD \cite{namd}, LAMMPS \cite{plimpton:jcop1995}, Tinker \cite{tinker8}, and the quantum chemistry code CP2K \cite{cp2knew}. Important work still remains in making the workflow easy enough for widespread use. 

\begin{acknowledgement} 

We thank our students and colleagues whose works are cited below for enriching our understanding of the theory. We gratefully acknowledge computing support from National Energy Research Scientific Computing Center (NERSC), a U.S. Department of Energy Office of Science User Facility located at Lawrence Berkeley National Laboratory, operated under Contract No. DE-AC02-05CH11231.  We gratefully acknowledge the Texas Advanced Computing Center (TACC) at The University of Texas at Austin for providing HPC resources that have contributed to the research results reported within this paper. 
\end{acknowledgement}

\newpage 


\providecommand{\latin}[1]{#1}
\makeatletter
\providecommand{\doi}
  {\begingroup\let\do\@makeother\dospecials
  \catcode`\{=1 \catcode`\}=2 \doi@aux}
\providecommand{\doi@aux}[1]{\endgroup\texttt{#1}}
\makeatother
\providecommand*\mcitethebibliography{\thebibliography}
\csname @ifundefined\endcsname{endmcitethebibliography}
  {\let\endmcitethebibliography\endthebibliography}{}
\begin{mcitethebibliography}{76}
\providecommand*\natexlab[1]{#1}
\providecommand*\mciteSetBstSublistMode[1]{}
\providecommand*\mciteSetBstMaxWidthForm[2]{}
\providecommand*\mciteBstWouldAddEndPuncttrue
  {\def\EndOfBibitem{\unskip.}}
\providecommand*\mciteBstWouldAddEndPunctfalse
  {\let\EndOfBibitem\relax}
\providecommand*\mciteSetBstMidEndSepPunct[3]{}
\providecommand*\mciteSetBstSublistLabelBeginEnd[3]{}
\providecommand*\EndOfBibitem{}
\mciteSetBstSublistMode{f}
\mciteSetBstMaxWidthForm{subitem}{(\alph{mcitesubitemcount})}
\mciteSetBstSublistLabelBeginEnd
  {\mcitemaxwidthsubitemform\space}
  {\relax}
  {\relax}

\bibitem[Paulaitis and Pratt(2002)Paulaitis, and Pratt]{lrp:apc02}
Paulaitis,~M.~E.; Pratt,~L.~R. Hydration Theory for Molecular Biophysics.
  \emph{Adv. Prot. Chem.} \textbf{2002}, \emph{62}, 283--310\relax
\mciteBstWouldAddEndPuncttrue
\mciteSetBstMidEndSepPunct{\mcitedefaultmidpunct}
{\mcitedefaultendpunct}{\mcitedefaultseppunct}\relax
\EndOfBibitem
\bibitem[Beck \latin{et~al.}(2006)Beck, Paulaitis, and Pratt]{lrp:book}
Beck,~T.~L.; Paulaitis,~M.~E.; Pratt,~L.~R. \emph{The Potential Distribution
  Theorem and Models of Molecular Solutions}; Cambridge University Press:
  Cambridge, UK, 2006\relax
\mciteBstWouldAddEndPuncttrue
\mciteSetBstMidEndSepPunct{\mcitedefaultmidpunct}
{\mcitedefaultendpunct}{\mcitedefaultseppunct}\relax
\EndOfBibitem
\bibitem[Pratt and Asthagiri(2007)Pratt, and Asthagiri]{lrp:cpms}
Pratt,~L.~R.; Asthagiri,~D. In \emph{Free Energy Calculations: {Theory} and
  Applications in Chemistry and Biology}; Chipot,~C., Pohorille,~A., Eds.;
  Springer series in {Chemical Physics}; Springer: Berlin, DE, 2007; Vol.~86;
  Chapter 9, pp 323--351\relax
\mciteBstWouldAddEndPuncttrue
\mciteSetBstMidEndSepPunct{\mcitedefaultmidpunct}
{\mcitedefaultendpunct}{\mcitedefaultseppunct}\relax
\EndOfBibitem
\bibitem[Asthagiri \latin{et~al.}(2010)Asthagiri, Dixit, Merchant, Paulaitis,
  Pratt, Rempe, and Varma]{asthagiri:cpl10}
Asthagiri,~D.; Dixit,~P.~D.; Merchant,~S.; Paulaitis,~M.~E.; Pratt,~L.~R.;
  Rempe,~S.~B.; Varma,~S. Ion Selectivity from Local Configurations of Ligands
  in Solution and Ion Channels. \emph{Chem. Phys. Lett.} \textbf{2010},
  \emph{485}, 1--7\relax
\mciteBstWouldAddEndPuncttrue
\mciteSetBstMidEndSepPunct{\mcitedefaultmidpunct}
{\mcitedefaultendpunct}{\mcitedefaultseppunct}\relax
\EndOfBibitem
\bibitem[Martin \latin{et~al.}(1998)Martin, Hay, and Pratt]{Martin:jpca98}
Martin,~R.~L.; Hay,~P.~J.; Pratt,~L.~R. {Hydrolysis of Ferric Ion in Water and
  Conformational Equilibrium}. \emph{J. Phys. Chem. A} \textbf{1998},
  \emph{102}, 3565--3573\relax
\mciteBstWouldAddEndPuncttrue
\mciteSetBstMidEndSepPunct{\mcitedefaultmidpunct}
{\mcitedefaultendpunct}{\mcitedefaultseppunct}\relax
\EndOfBibitem
\bibitem[Rempe \latin{et~al.}(2000)Rempe, Pratt, Hummer, Kress, Martin, and
  Redondo]{rempe:lijacs}
Rempe,~S.~B.; Pratt,~L.~R.; Hummer,~G.; Kress,~J.~D.; Martin,~R.~L.;
  Redondo,~A. The Hydration Number of {Li}$^+$ in Liquid Water. \emph{J. Am.
  Chem. Soc.} \textbf{2000}, \emph{122}, 966--967\relax
\mciteBstWouldAddEndPuncttrue
\mciteSetBstMidEndSepPunct{\mcitedefaultmidpunct}
{\mcitedefaultendpunct}{\mcitedefaultseppunct}\relax
\EndOfBibitem
\bibitem[Friedman(1985)]{friedman1985hydration}
Friedman,~H. Hydration Complexes --- Some Firm Results and Some Pressing
  Questions. \emph{Chemica Scripta} \textbf{1985}, \emph{25}, 42--48\relax
\mciteBstWouldAddEndPuncttrue
\mciteSetBstMidEndSepPunct{\mcitedefaultmidpunct}
{\mcitedefaultendpunct}{\mcitedefaultseppunct}\relax
\EndOfBibitem
\bibitem[Howell and Neilson(1996)Howell, and Neilson]{howell1996hydration}
Howell,~I.; Neilson,~G. Hydration in Concentrated Aqueous Solution. \emph{J.
  Phys.: Cond. Matt.} \textbf{1996}, \emph{8}, 4455\relax
\mciteBstWouldAddEndPuncttrue
\mciteSetBstMidEndSepPunct{\mcitedefaultmidpunct}
{\mcitedefaultendpunct}{\mcitedefaultseppunct}\relax
\EndOfBibitem
\bibitem[Varma and Rempe(2006)Varma, and Rempe]{rempe:bc06}
Varma,~S.; Rempe,~S.~B. Coordination Numbers of Alkali Metal Ions in Aqueous
  Solutions. \emph{Biophys. Chem.} \textbf{2006}, \emph{124}, 192--199\relax
\mciteBstWouldAddEndPuncttrue
\mciteSetBstMidEndSepPunct{\mcitedefaultmidpunct}
{\mcitedefaultendpunct}{\mcitedefaultseppunct}\relax
\EndOfBibitem
\bibitem[Mason \latin{et~al.}(2015)Mason, Ansell, Neilson, and
  Rempe]{mason2015neutron}
Mason,~P.; Ansell,~S.; Neilson,~G.; Rempe,~S. Neutron Scattering Studies of the
  Hydration Structure of {Li}$^+$. \emph{J. Phys. Chem. B} \textbf{2015},
  \emph{119}, 2003--2009\relax
\mciteBstWouldAddEndPuncttrue
\mciteSetBstMidEndSepPunct{\mcitedefaultmidpunct}
{\mcitedefaultendpunct}{\mcitedefaultseppunct}\relax
\EndOfBibitem
\bibitem[Asthagiri \latin{et~al.}(2004)Asthagiri, Pratt, Paulaitis, and
  Rempe]{Asthagiri:jacs04}
Asthagiri,~D.; Pratt,~L.~R.; Paulaitis,~M.~E.; Rempe,~S.~B. Hydration Structure
  and Free Energy of Biomolecularly Specific Aqueous Dications, Including
  {Zn}$^{2+}$ and First Transition Row Metals. \emph{J. Am. Chem. Soc.}
  \textbf{2004}, \emph{126}, 1285--1289\relax
\mciteBstWouldAddEndPuncttrue
\mciteSetBstMidEndSepPunct{\mcitedefaultmidpunct}
{\mcitedefaultendpunct}{\mcitedefaultseppunct}\relax
\EndOfBibitem
\bibitem[Asthagiri and Pratt(2003)Asthagiri, and Pratt]{Asthagiri:becpl}
Asthagiri,~D.; Pratt,~L.~R. Quasi-chemical Study of {Be}$^{2+}$(aq) Speciation.
  \emph{Chem. Phys. Lett.} \textbf{2003}, \emph{371}, 613--619\relax
\mciteBstWouldAddEndPuncttrue
\mciteSetBstMidEndSepPunct{\mcitedefaultmidpunct}
{\mcitedefaultendpunct}{\mcitedefaultseppunct}\relax
\EndOfBibitem
\bibitem[Asthagiri \latin{et~al.}(2003)Asthagiri, Pratt, Kress, and
  Gomez]{Asthagiri:hocpl}
Asthagiri,~D.; Pratt,~L.~R.; Kress,~J.~D.; Gomez,~M.~A. The Hydration State of
  {HO}$^-$(aq). \emph{Chem. Phys. Lett.} \textbf{2003}, \emph{380},
  530--535\relax
\mciteBstWouldAddEndPuncttrue
\mciteSetBstMidEndSepPunct{\mcitedefaultmidpunct}
{\mcitedefaultendpunct}{\mcitedefaultseppunct}\relax
\EndOfBibitem
\bibitem[Grabowski \latin{et~al.}(2002)Grabowski, Riccardi, Gomez, Asthagiri,
  and Pratt]{grabowski:jpca02}
Grabowski,~P.; Riccardi,~D.; Gomez,~M.~A.; Asthagiri,~D.; Pratt,~L.~R.
  Quasi-chemical Theory and the Standard Free Energy of {H}$^+$(aq). \emph{J.
  Phys. Chem. A} \textbf{2002}, \emph{106}, 9145--9148\relax
\mciteBstWouldAddEndPuncttrue
\mciteSetBstMidEndSepPunct{\mcitedefaultmidpunct}
{\mcitedefaultendpunct}{\mcitedefaultseppunct}\relax
\EndOfBibitem
\bibitem[Paliwal \latin{et~al.}(2006)Paliwal, Asthagiri, Pratt, Ashbaugh, and
  Paulaitis]{paliwal:jcp06}
Paliwal,~A.; Asthagiri,~D.; Pratt,~L.~R.; Ashbaugh,~H.~S.; Paulaitis,~M.~E. An
  Analysis of Molecular Packing and Chemical Association in Liquid Water Using
  Quasichemical Theory. \emph{J. Chem. Phys.} \textbf{2006}, \emph{124},
  224502\relax
\mciteBstWouldAddEndPuncttrue
\mciteSetBstMidEndSepPunct{\mcitedefaultmidpunct}
{\mcitedefaultendpunct}{\mcitedefaultseppunct}\relax
\EndOfBibitem
\bibitem[Shah \latin{et~al.}(2007)Shah, Asthagiri, Pratt, and
  Paulaitis]{shah:jcp07}
Shah,~J.~K.; Asthagiri,~D.; Pratt,~L.~R.; Paulaitis,~M.~E. Balancing Local
  Order and Long-Ranged Interactions in the Molecular Theory of Liquid Water.
  \emph{J. Chem. Phys.} \textbf{2007}, \emph{127}, 144508\relax
\mciteBstWouldAddEndPuncttrue
\mciteSetBstMidEndSepPunct{\mcitedefaultmidpunct}
{\mcitedefaultendpunct}{\mcitedefaultseppunct}\relax
\EndOfBibitem
\bibitem[Chempath \latin{et~al.}(2009)Chempath, Pratt, and
  Paulaitis]{lrp:softcutoff}
Chempath,~S.; Pratt,~L.~R.; Paulaitis,~M.~E. Quasi-chemical Theory with a Soft
  Cutoff. \emph{J. Chem. Phys.} \textbf{2009}, \emph{130}, 054113\relax
\mciteBstWouldAddEndPuncttrue
\mciteSetBstMidEndSepPunct{\mcitedefaultmidpunct}
{\mcitedefaultendpunct}{\mcitedefaultseppunct}\relax
\EndOfBibitem
\bibitem[Weber \latin{et~al.}(2010)Weber, Merchant, Dixit, and
  Asthagiri]{weber:jcp10a}
Weber,~V.; Merchant,~S.; Dixit,~P.~D.; Asthagiri,~D. Molecular Packing and
  Chemical Association in Liquid Water Simulated Using \emph{Ab Initio} Hybrid
  {Monte Carlo} and Different Exchange-Correlation Functionals. \emph{J. Chem.
  Phys.} \textbf{2010}, \emph{132}, 204509\relax
\mciteBstWouldAddEndPuncttrue
\mciteSetBstMidEndSepPunct{\mcitedefaultmidpunct}
{\mcitedefaultendpunct}{\mcitedefaultseppunct}\relax
\EndOfBibitem
\bibitem[Weber and Asthagiri(2010)Weber, and Asthagiri]{weber:jcp10b}
Weber,~V.; Asthagiri,~D. Thermodynamics of Water Modeled Using \emph{Ab Initio}
  Simulations. \emph{J. Chem. Phys.} \textbf{2010}, \emph{133}, 141101\relax
\mciteBstWouldAddEndPuncttrue
\mciteSetBstMidEndSepPunct{\mcitedefaultmidpunct}
{\mcitedefaultendpunct}{\mcitedefaultseppunct}\relax
\EndOfBibitem
\bibitem[Chaudhari \latin{et~al.}(2014)Chaudhari, Pratt, and
  Paulaitis]{chaudhari2014concentration}
Chaudhari,~M.~I.; Pratt,~L.~R.; Paulaitis,~M.~E. Concentration Dependence of
  the Flory-Huggins Interaction Parameter in Aqueous Solutions of Capped PEO
  Chains. \emph{J. Chem. Phys.} \textbf{2014}, \emph{141}, 244908\relax
\mciteBstWouldAddEndPuncttrue
\mciteSetBstMidEndSepPunct{\mcitedefaultmidpunct}
{\mcitedefaultendpunct}{\mcitedefaultseppunct}\relax
\EndOfBibitem
\bibitem[Weber and Asthagiri(2012)Weber, and Asthagiri]{Weber:jctc12}
Weber,~V.; Asthagiri,~D. Regularizing Binding Energy Distributions and the
  Hydration Free Energy of Protein {Cytochrome C} from All-Atom Simulations.
  \emph{J. Chem. Theory Comput.} \textbf{2012}, \emph{8}, 3409--3415\relax
\mciteBstWouldAddEndPuncttrue
\mciteSetBstMidEndSepPunct{\mcitedefaultmidpunct}
{\mcitedefaultendpunct}{\mcitedefaultseppunct}\relax
\EndOfBibitem
\bibitem[Utiramerur and Paulaitis(2010)Utiramerur, and
  Paulaitis]{paulaitis:corr10}
Utiramerur,~S.; Paulaitis,~M.~E. {Cooperative Hydrophobic/Hydrophilic
  Interactions in the Hydration of Dimethyl Ether}. \emph{J. Chem. Phys.}
  \textbf{2010}, \emph{132}, 155102\relax
\mciteBstWouldAddEndPuncttrue
\mciteSetBstMidEndSepPunct{\mcitedefaultmidpunct}
{\mcitedefaultendpunct}{\mcitedefaultseppunct}\relax
\EndOfBibitem
\bibitem[Tomar \latin{et~al.}(2013)Tomar, Weber, and Asthagiri]{tomar:bj2013}
Tomar,~D.~S.; Weber,~V.; Asthagiri,~D. Solvation Free Energy of the Peptide
  Group: Its Model Dependence and Implications for the Additive Transfer Free
  Energy Model. \emph{Biophys. J.} \textbf{2013}, \emph{105}, 1482--1490\relax
\mciteBstWouldAddEndPuncttrue
\mciteSetBstMidEndSepPunct{\mcitedefaultmidpunct}
{\mcitedefaultendpunct}{\mcitedefaultseppunct}\relax
\EndOfBibitem
\bibitem[Tomar \latin{et~al.}(2014)Tomar, Weber, Pettitt, and
  Asthagiri]{tomar:jpcb14}
Tomar,~D.~S.; Weber,~V.; Pettitt,~B.~M.; Asthagiri,~D. Conditional Solvation
  Thermodynamics of Isoleucine in Model Peptides and the Limitations of the
  Group-Transfer Model. \emph{J. Phys. Chem. B} \textbf{2014}, \emph{118},
  4080--4087\relax
\mciteBstWouldAddEndPuncttrue
\mciteSetBstMidEndSepPunct{\mcitedefaultmidpunct}
{\mcitedefaultendpunct}{\mcitedefaultseppunct}\relax
\EndOfBibitem
\bibitem[Tomar \latin{et~al.}(2018)Tomar, Ramesh, and
  Asthagiri]{tomar:gdmjcp18}
Tomar,~D.~S.; Ramesh,~N.; Asthagiri,~D. {Solvophobic and Solvophilic
  Contributions in the Water-To-Aqueous Guanidinium Chloride Transfer Free
  Energy of Model Peptides}. \emph{J. Chem. Phys.} \textbf{2018}, \emph{148},
  222822\relax
\mciteBstWouldAddEndPuncttrue
\mciteSetBstMidEndSepPunct{\mcitedefaultmidpunct}
{\mcitedefaultendpunct}{\mcitedefaultseppunct}\relax
\EndOfBibitem
\bibitem[Tomar \latin{et~al.}(2016)Tomar, Weber, Pettitt, and
  Asthagiri]{tomar:jpcb16}
Tomar,~D.~S.; Weber,~W.; Pettitt,~M.~B.; Asthagiri,~D. {Importance of
  Hydrophilic Hydration and Intramolecular Interactions in the Thermodynamics
  of Helix-Coil Transition and Helix-Helix Assembly in a Deca-Alanine Peptide.}
  \emph{J. Phys. Chem. B} \textbf{2016}, \emph{120}, 69--76\relax
\mciteBstWouldAddEndPuncttrue
\mciteSetBstMidEndSepPunct{\mcitedefaultmidpunct}
{\mcitedefaultendpunct}{\mcitedefaultseppunct}\relax
\EndOfBibitem
\bibitem[Asthagiri \latin{et~al.}(2017)Asthagiri, Karandur, Tomar, and
  Pettitt]{asthagiri:gly15}
Asthagiri,~D.; Karandur,~D.; Tomar,~D.~S.; Pettitt,~B.~M. {Intramolecular
  Interactions Overcome Hydration to Drive the Collapse Transition of
  Gly$_{15}$}. \emph{J. Phys. Chem. B} \textbf{2017}, \emph{121},
  8078--8084\relax
\mciteBstWouldAddEndPuncttrue
\mciteSetBstMidEndSepPunct{\mcitedefaultmidpunct}
{\mcitedefaultendpunct}{\mcitedefaultseppunct}\relax
\EndOfBibitem
\bibitem[Tomar \latin{et~al.}(2020)Tomar, Paulaitis, Pratt, and
  Asthagiri]{tomar:jpcl20}
Tomar,~D.~S.; Paulaitis,~M.~E.; Pratt,~L.~R.; Asthagiri,~D.~N. {Hydrophilic
  Interactions Dominate the Inverse Temperature Dependence of Polypeptide
  Hydration Free Energies Attributed to Hydrophobicity}. \emph{J. Phys. Chem.
  Lett.} \textbf{2020}, \emph{11}, 9965--9970\relax
\mciteBstWouldAddEndPuncttrue
\mciteSetBstMidEndSepPunct{\mcitedefaultmidpunct}
{\mcitedefaultendpunct}{\mcitedefaultseppunct}\relax
\EndOfBibitem
\bibitem[Widom(1982)]{widom:jpc82}
Widom,~B. Potential-distribution Theory and the Statistical Mechanics of
  Fluids. \emph{J. Phys. Chem.} \textbf{1982}, \emph{86}, 869--872\relax
\mciteBstWouldAddEndPuncttrue
\mciteSetBstMidEndSepPunct{\mcitedefaultmidpunct}
{\mcitedefaultendpunct}{\mcitedefaultseppunct}\relax
\EndOfBibitem
\bibitem[Pratt and Rempe(1999)Pratt, and Rempe]{lrp:ES99}
Pratt,~L.~R.; Rempe,~S.~B. In \emph{Simulation and Theory of Electrostatic
  Interactions in Solution. {Computational} Chemistry, Biophysics, and Aqueous
  Solutions}; Pratt,~L.~R., Hummer,~G., Eds.; AIP Conference Proceedings;
  American Institute of Physics: Melville, NY, 1999; Vol. 492; pp
  172--201\relax
\mciteBstWouldAddEndPuncttrue
\mciteSetBstMidEndSepPunct{\mcitedefaultmidpunct}
{\mcitedefaultendpunct}{\mcitedefaultseppunct}\relax
\EndOfBibitem
\bibitem[Zhang \latin{et~al.}(2014)Zhang, You, and Pratt]{Zhang:jpcb14}
Zhang,~W.; You,~X.; Pratt,~L.~R. {Multiscale Theory in the Molecular Simulation
  of Electrolyte Solutions}. \emph{J. Phys. Chem. B} \textbf{2014}, \emph{118},
  7730--7738\relax
\mciteBstWouldAddEndPuncttrue
\mciteSetBstMidEndSepPunct{\mcitedefaultmidpunct}
{\mcitedefaultendpunct}{\mcitedefaultseppunct}\relax
\EndOfBibitem
\bibitem[Vafaei \latin{et~al.}(2014)Vafaei, Tomberli, and Gray]{Vafaei:jcp14}
Vafaei,~S.; Tomberli,~B.; Gray,~C.~G. {McMillan-Mayer Theory of Solutions
  Revisited: Simplifications and Extensions}. \emph{J. Chem. Phys.}
  \textbf{2014}, \emph{141}, 154501\relax
\mciteBstWouldAddEndPuncttrue
\mciteSetBstMidEndSepPunct{\mcitedefaultmidpunct}
{\mcitedefaultendpunct}{\mcitedefaultseppunct}\relax
\EndOfBibitem
\bibitem[Chipot and Pohorille(2007)Chipot, and Pohorille]{chipotpohorille}
Chipot,~C., Pohorille,~A., Eds. \emph{Free Energy Calculations: Theory and
  Applications in Chemistry and Biology}; Springer Series in Chemical Physics;
  Springer: Berlin, 2007; Vol.~86\relax
\mciteBstWouldAddEndPuncttrue
\mciteSetBstMidEndSepPunct{\mcitedefaultmidpunct}
{\mcitedefaultendpunct}{\mcitedefaultseppunct}\relax
\EndOfBibitem
\bibitem[Merchant(2011)]{safirphd}
Merchant,~S. Regularizing Free Energy Calculations to Study Ion Specific
  Effects in Biology. Ph.D.\ thesis, Johns Hopkins University, Baltimore,
  2011\relax
\mciteBstWouldAddEndPuncttrue
\mciteSetBstMidEndSepPunct{\mcitedefaultmidpunct}
{\mcitedefaultendpunct}{\mcitedefaultseppunct}\relax
\EndOfBibitem
\bibitem[Pratt and Ashbaugh(2003)Pratt, and Ashbaugh]{lrp:hspre}
Pratt,~L.~R.; Ashbaugh,~H.~S. Self-Consistent Molecular Field Theory for
  Packing in Classical Liquids. \emph{Phys. Rev. E} \textbf{2003}, \emph{68},
  021505\relax
\mciteBstWouldAddEndPuncttrue
\mciteSetBstMidEndSepPunct{\mcitedefaultmidpunct}
{\mcitedefaultendpunct}{\mcitedefaultseppunct}\relax
\EndOfBibitem
\bibitem[Merchant \latin{et~al.}(2011)Merchant, Dixit, Dean, and
  Asthagiri]{merchant:jcp11b}
Merchant,~S.; Dixit,~P.~D.; Dean,~K.~R.; Asthagiri,~D. Ion-Water Clusters, Bulk
  Medium Effects, and Ion Hydration. \emph{J. Chem. Phys.} \textbf{2011},
  \emph{135}, 054505\relax
\mciteBstWouldAddEndPuncttrue
\mciteSetBstMidEndSepPunct{\mcitedefaultmidpunct}
{\mcitedefaultendpunct}{\mcitedefaultseppunct}\relax
\EndOfBibitem
\bibitem[Merchant and Asthagiri(2009)Merchant, and Asthagiri]{merchant:jcp09}
Merchant,~S.; Asthagiri,~D. Thermodynamically Dominant Hydration Structures of
  Aqueous Ions. \emph{J. Chem. Phys.} \textbf{2009}, \emph{130}, 195102\relax
\mciteBstWouldAddEndPuncttrue
\mciteSetBstMidEndSepPunct{\mcitedefaultmidpunct}
{\mcitedefaultendpunct}{\mcitedefaultseppunct}\relax
\EndOfBibitem
\bibitem[Weber \latin{et~al.}(2011)Weber, Merchant, and Asthagiri]{weber:jcp11}
Weber,~V.; Merchant,~S.; Asthagiri,~D. Regularizing Binding Energy
  Distributions and Thermodynamics of Hydration: {Theory} and Application to
  Water Modeled with Classical and \emph{Ab Initio} Simulations. \emph{J. Chem.
  Phys.} \textbf{2011}, \emph{135}, 181101\relax
\mciteBstWouldAddEndPuncttrue
\mciteSetBstMidEndSepPunct{\mcitedefaultmidpunct}
{\mcitedefaultendpunct}{\mcitedefaultseppunct}\relax
\EndOfBibitem
\bibitem[Asthagiri and Tomar(2020)Asthagiri, and Tomar]{asthagiri:jpcb20a}
Asthagiri,~D.; Tomar,~D.~S. System Size Dependence of Hydration-Shell Occupancy
  and Its Implications for Assessing the Hydrophobic and Hydrophilic
  Contributions to Hydration. \emph{J. Phys. Chem. B} \textbf{2020},
  \emph{124}, 798--806\relax
\mciteBstWouldAddEndPuncttrue
\mciteSetBstMidEndSepPunct{\mcitedefaultmidpunct}
{\mcitedefaultendpunct}{\mcitedefaultseppunct}\relax
\EndOfBibitem
\bibitem[Asthagiri \latin{et~al.}(2008)Asthagiri, Merchant, and
  Pratt]{asthagiri:jcp2008}
Asthagiri,~D.; Merchant,~S.; Pratt,~L.~R. Role of Attractive Methane-Water
  Interactions in the Potential of Mean Force Between Methane Molecules in
  Water. \emph{J. Chem. Phys.} \textbf{2008}, \emph{128}, 244512\relax
\mciteBstWouldAddEndPuncttrue
\mciteSetBstMidEndSepPunct{\mcitedefaultmidpunct}
{\mcitedefaultendpunct}{\mcitedefaultseppunct}\relax
\EndOfBibitem
\bibitem[Rempe and Pratt(2001)Rempe, and Pratt]{rempe:nafpe}
Rempe,~S.~B.; Pratt,~L.~R. The Hydration Number of {Na}$^+$ in Liquid Water.
  \emph{Fluid Phase Equil.} \textbf{2001}, \emph{183-184}, 121--132\relax
\mciteBstWouldAddEndPuncttrue
\mciteSetBstMidEndSepPunct{\mcitedefaultmidpunct}
{\mcitedefaultendpunct}{\mcitedefaultseppunct}\relax
\EndOfBibitem
\bibitem[Ashbaugh \latin{et~al.}(2003)Ashbaugh, Asthagiri, Pratt, and
  Rempe]{Ashbaugh:kr03}
Ashbaugh,~H.~S.; Asthagiri,~D.; Pratt,~L.~R.; Rempe,~S.~B. {Hydration of
  Krypton and Consideration of Clathrate Models of Hydrophobic Effects from the
  Perspective of Quasi-Chemical Theory}. \emph{Biophys. Chem.} \textbf{2003},
  \emph{105}, 323--338\relax
\mciteBstWouldAddEndPuncttrue
\mciteSetBstMidEndSepPunct{\mcitedefaultmidpunct}
{\mcitedefaultendpunct}{\mcitedefaultseppunct}\relax
\EndOfBibitem
\bibitem[Asthagiri \latin{et~al.}(2003)Asthagiri, Pratt, and
  Ashbaugh]{asthagiri:jcp03}
Asthagiri,~D.; Pratt,~L.~R.; Ashbaugh,~H.~S. Absolute Hydration Free Energies
  of Ions, Ion-Water Clusters, and Quasichemical Theory. \emph{J. Chem. Phys.}
  \textbf{2003}, \emph{119}, 2702--2708\relax
\mciteBstWouldAddEndPuncttrue
\mciteSetBstMidEndSepPunct{\mcitedefaultmidpunct}
{\mcitedefaultendpunct}{\mcitedefaultseppunct}\relax
\EndOfBibitem
\bibitem[Asthagiri \latin{et~al.}(2004)Asthagiri, Pratt, Kress, and
  Gomez]{Asthagiri:pnas04}
Asthagiri,~D.; Pratt,~L.~R.; Kress,~J.~D.; Gomez,~M.~A. Hydration and Mobility
  of {HO}$^-$(aq). \emph{Proc. Natl. Acad. Sci. USA} \textbf{2004}, \emph{101},
  7229--7233\relax
\mciteBstWouldAddEndPuncttrue
\mciteSetBstMidEndSepPunct{\mcitedefaultmidpunct}
{\mcitedefaultendpunct}{\mcitedefaultseppunct}\relax
\EndOfBibitem
\bibitem[Asthagiri \latin{et~al.}(2005)Asthagiri, Pratt, and
  Kress]{Asthagiri:pnas05}
Asthagiri,~D.; Pratt,~L.~R.; Kress,~J.~D. {\em Ab Initio} Molecular Dynamics
  and Quasichemical Study of {H}$^+$(aq). \emph{Proc. Natl. Acad. Sci. USA}
  \textbf{2005}, \emph{102}, 6704--6708\relax
\mciteBstWouldAddEndPuncttrue
\mciteSetBstMidEndSepPunct{\mcitedefaultmidpunct}
{\mcitedefaultendpunct}{\mcitedefaultseppunct}\relax
\EndOfBibitem
\bibitem[Rempe \latin{et~al.}(2004)Rempe, Asthagiri, and Pratt]{rempe:kpccp}
Rempe,~S.~B.; Asthagiri,~D.; Pratt,~L.~R. Inner Shell Definition and Absolute
  Hydration Free Energy of {K}$^+$(aq) on the Basis of Quasi-Chemical Theory
  and {\em Ab Initio\/} Molecular Dynamics. \emph{Phys. Chem. Chem. Phys.}
  \textbf{2004}, \emph{6}, 1966--1969\relax
\mciteBstWouldAddEndPuncttrue
\mciteSetBstMidEndSepPunct{\mcitedefaultmidpunct}
{\mcitedefaultendpunct}{\mcitedefaultseppunct}\relax
\EndOfBibitem
\bibitem[Rogers and Beck(2008)Rogers, and Beck]{beck:jcp08}
Rogers,~D.~M.; Beck,~T.~L. Modeling Molecular and Ionic Absolute Solvation Free
  Energies with Quasichemical Theory Bounds. \emph{J. Chem. Phys.}
  \textbf{2008}, \emph{129}, 134505\relax
\mciteBstWouldAddEndPuncttrue
\mciteSetBstMidEndSepPunct{\mcitedefaultmidpunct}
{\mcitedefaultendpunct}{\mcitedefaultseppunct}\relax
\EndOfBibitem
\bibitem[Rogers and Beck(2010)Rogers, and Beck]{beck:jcp10}
Rogers,~D.~M.; Beck,~T.~L. Quasichemical and Structural Analysis of Polarizable
  Anion Hydration. \emph{J. Chem. Phys.} \textbf{2010}, \emph{132},
  014505\relax
\mciteBstWouldAddEndPuncttrue
\mciteSetBstMidEndSepPunct{\mcitedefaultmidpunct}
{\mcitedefaultendpunct}{\mcitedefaultseppunct}\relax
\EndOfBibitem
\bibitem[Beck(2011)]{beck:jpcb11}
Beck,~T.~L. A Local Entropic Signature of Specific Ion Hydration. \emph{J.
  Phys. Chem. B} \textbf{2011}, \emph{115}, 9776--9781\relax
\mciteBstWouldAddEndPuncttrue
\mciteSetBstMidEndSepPunct{\mcitedefaultmidpunct}
{\mcitedefaultendpunct}{\mcitedefaultseppunct}\relax
\EndOfBibitem
\bibitem[Asthagiri \latin{et~al.}(2003)Asthagiri, Pratt, and
  Kress]{Asthagiri:pre03}
Asthagiri,~D.; Pratt,~L.~R.; Kress,~J.~D. Free Energy of Liquid Water on the
  Basis of Quasichemical Theory and {\em Ab Initio} Molecular Dynamics.
  \emph{Phys. Rev. E} \textbf{2003}, \emph{68}, 041505\relax
\mciteBstWouldAddEndPuncttrue
\mciteSetBstMidEndSepPunct{\mcitedefaultmidpunct}
{\mcitedefaultendpunct}{\mcitedefaultseppunct}\relax
\EndOfBibitem
\bibitem[Chempath and Pratt(2009)Chempath, and Pratt]{lrp:jpcb09}
Chempath,~S.; Pratt,~L.~R. Distribution of Binding Energies of a Water Molecule
  in the Water Liquid-Vapor Interface. \emph{J. Phys. Chem. B} \textbf{2009},
  \emph{113}, 4147--4151\relax
\mciteBstWouldAddEndPuncttrue
\mciteSetBstMidEndSepPunct{\mcitedefaultmidpunct}
{\mcitedefaultendpunct}{\mcitedefaultseppunct}\relax
\EndOfBibitem
\bibitem[Merchant \latin{et~al.}(2010)Merchant, Shah, and
  Asthagiri]{merchant:jcp11a}
Merchant,~S.; Shah,~J.~K.; Asthagiri,~D. Water Coordination Structures and the
  Excess Free Energy of the Liquid. \emph{J. Chem. Phys.} \textbf{2010},
  \emph{134}, 124514\relax
\mciteBstWouldAddEndPuncttrue
\mciteSetBstMidEndSepPunct{\mcitedefaultmidpunct}
{\mcitedefaultendpunct}{\mcitedefaultseppunct}\relax
\EndOfBibitem
\bibitem[Asthagiri \latin{et~al.}(2007)Asthagiri, Ashbaugh, Piryatinski,
  Paulaitis, and Pratt]{asthagiri2007non}
Asthagiri,~D.; Ashbaugh,~H.~S.; Piryatinski,~A.; Paulaitis,~M.~E.; Pratt,~L.~R.
  Non-van der {W}aals Treatment of the Hydrophobic Solubilities of {CF}$_4$.
  \emph{J. Am. Chem. Soc.} \textbf{2007}, \emph{129}, 10133--10140\relax
\mciteBstWouldAddEndPuncttrue
\mciteSetBstMidEndSepPunct{\mcitedefaultmidpunct}
{\mcitedefaultendpunct}{\mcitedefaultseppunct}\relax
\EndOfBibitem
\bibitem[Utiramerur and Paulaitis(2021)Utiramerur, and
  Paulaitis]{paulaitis:corr21}
Utiramerur,~S.; Paulaitis,~M.~E. {Analysis of Cooperativity and Group
  Additivity in the Hydration of 1, 2-dimethoxyethane}. \emph{J. Phys. Chem. B}
  \textbf{2021}, \emph{125}, 1660--1666\relax
\mciteBstWouldAddEndPuncttrue
\mciteSetBstMidEndSepPunct{\mcitedefaultmidpunct}
{\mcitedefaultendpunct}{\mcitedefaultseppunct}\relax
\EndOfBibitem
\bibitem[Pratt \latin{et~al.}(2001)Pratt, LaViolette, Gomez, and
  Gentile]{lrp:jpcb01}
Pratt,~L.~R.; LaViolette,~R.~A.; Gomez,~M.~A.; Gentile,~M.~E. Quasi-Chemical
  Theory for the Statistical Thermodynamics of the Hard-Sphere Fluid. \emph{J.
  Phys. Chem. B} \textbf{2001}, \emph{105}, 11662--11668\relax
\mciteBstWouldAddEndPuncttrue
\mciteSetBstMidEndSepPunct{\mcitedefaultmidpunct}
{\mcitedefaultendpunct}{\mcitedefaultseppunct}\relax
\EndOfBibitem
\bibitem[Tam \latin{et~al.}(2012)Tam, Asthagiri, and Paulaitis]{Tam:2012jo}
Tam,~H.~H.; Asthagiri,~D.; Paulaitis,~M.~E. {Coordination State Probabilities
  and the Solvation Free Energy of Zn$^{2+}$ in Aqueous Methanol Solutions}.
  \emph{J. Chem. Phys.} \textbf{2012}, \emph{137}, 164504\relax
\mciteBstWouldAddEndPuncttrue
\mciteSetBstMidEndSepPunct{\mcitedefaultmidpunct}
{\mcitedefaultendpunct}{\mcitedefaultseppunct}\relax
\EndOfBibitem
\bibitem[Bansal \latin{et~al.}(2017)Bansal, Chapman, and
  Asthagiri]{Bansal:2017iv}
Bansal,~A.; Chapman,~W.~G.; Asthagiri,~D. {Quasichemical Theory and the
  Description of Associating Fluids Relative to a Reference: Multiple Bonding
  of a Single Site Solute}. \emph{J. Chem. Phys.} \textbf{2017}, \emph{147},
  124505\relax
\mciteBstWouldAddEndPuncttrue
\mciteSetBstMidEndSepPunct{\mcitedefaultmidpunct}
{\mcitedefaultendpunct}{\mcitedefaultseppunct}\relax
\EndOfBibitem
\bibitem[Remsing \latin{et~al.}(2018)Remsing, Xi, and Patel]{rxp}
Remsing,~R.~C.; Xi,~E.; Patel,~A.~J. Protein Hydration Thermodynamics: The
  Influence of Flexibility and Salt on Hydrophobin II Hydration. \emph{J. Phys.
  Chem. B} \textbf{2018}, \emph{122}, 3635--3646\relax
\mciteBstWouldAddEndPuncttrue
\mciteSetBstMidEndSepPunct{\mcitedefaultmidpunct}
{\mcitedefaultendpunct}{\mcitedefaultseppunct}\relax
\EndOfBibitem
\bibitem[Pratt and Chandler(1980)Pratt, and Chandler]{pratt1980effects}
Pratt,~L.~R.; Chandler,~D. Effects of Solute--Solvent Attractive Forces on
  Hydrophobic Correlations. \emph{J. Chem. Phys.} \textbf{1980}, \emph{73},
  3434--3441\relax
\mciteBstWouldAddEndPuncttrue
\mciteSetBstMidEndSepPunct{\mcitedefaultmidpunct}
{\mcitedefaultendpunct}{\mcitedefaultseppunct}\relax
\EndOfBibitem
\bibitem[Rossky and Friedman(1980)Rossky, and Friedman]{rossky1980benzene}
Rossky,~P.~J.; Friedman,~H.~L. Benzene-Benzene Interaction in Aqueous Solution.
  \emph{J. Phys. Chem.} \textbf{1980}, \emph{84}, 587--589\relax
\mciteBstWouldAddEndPuncttrue
\mciteSetBstMidEndSepPunct{\mcitedefaultmidpunct}
{\mcitedefaultendpunct}{\mcitedefaultseppunct}\relax
\EndOfBibitem
\bibitem[Pratt(1985)]{pratt1985theory}
Pratt,~L.~R. Theory of Hydrophobic Effects. \emph{Ann. Rev. Phys. Chem.}
  \textbf{1985}, \emph{36}, 433--449\relax
\mciteBstWouldAddEndPuncttrue
\mciteSetBstMidEndSepPunct{\mcitedefaultmidpunct}
{\mcitedefaultendpunct}{\mcitedefaultseppunct}\relax
\EndOfBibitem
\bibitem[Chaudhari \latin{et~al.}(2015)Chaudhari, Rempe, Asthagiri, Tan, and
  Pratt]{chaudhari:jpcb15}
Chaudhari,~M.~I.; Rempe,~S.~R.; Asthagiri,~D.; Tan,~L.; Pratt,~L.~R. Molecular
  Theory and the Effects of Solute Attractive Forces on Hydrophobic
  Interactions. \emph{J. Phys. Chem. B} \textbf{2015}, \emph{120},
  1864--1870\relax
\mciteBstWouldAddEndPuncttrue
\mciteSetBstMidEndSepPunct{\mcitedefaultmidpunct}
{\mcitedefaultendpunct}{\mcitedefaultseppunct}\relax
\EndOfBibitem
\bibitem[Pratt \latin{et~al.}(2016)Pratt, Chaudhari, and
  Rempe]{pratt2016statistical}
Pratt,~L.~R.; Chaudhari,~M.~I.; Rempe,~S.~B. Statistical Analyses of
  Hydrophobic Interactions: A Mini-Review. \emph{J. Phys. Chem. B}
  \textbf{2016}, \emph{120}, 6455--6460\relax
\mciteBstWouldAddEndPuncttrue
\mciteSetBstMidEndSepPunct{\mcitedefaultmidpunct}
{\mcitedefaultendpunct}{\mcitedefaultseppunct}\relax
\EndOfBibitem
\bibitem[Gao \latin{et~al.}(2018)Gao, Tan, Chaudhari, Asthagiri, Pratt, Rempe,
  and Weeks]{gao2018role}
Gao,~A.; Tan,~L.; Chaudhari,~M.~I.; Asthagiri,~D.; Pratt,~L.~R.; Rempe,~S.~B.;
  Weeks,~J.~D. Role of Solute Attractive Forces in the Atomic-Scale Theory of
  Hydrophobic Effects. \emph{J. Phys. Chem. B} \textbf{2018}, \emph{122},
  6272--6276\relax
\mciteBstWouldAddEndPuncttrue
\mciteSetBstMidEndSepPunct{\mcitedefaultmidpunct}
{\mcitedefaultendpunct}{\mcitedefaultseppunct}\relax
\EndOfBibitem
\bibitem[Alexander \latin{et~al.}(2009)Alexander, He, Chen, Orban, and
  Bryan]{bryan:pnas09}
Alexander,~P.~A.; He,~Y.; Chen,~Y.; Orban,~J.; Bryan,~P.~N. A Minimal Sequence
  Code for Switching Protein Structure and Function. \emph{Proc. Natl. Acad.
  Sci. USA} \textbf{2009}, \emph{106}, 21149--21154\relax
\mciteBstWouldAddEndPuncttrue
\mciteSetBstMidEndSepPunct{\mcitedefaultmidpunct}
{\mcitedefaultendpunct}{\mcitedefaultseppunct}\relax
\EndOfBibitem
\bibitem[Garde \latin{et~al.}(1996)Garde, Hummer, Garcia, Paulaitis, and
  Pratt]{garde:prl96}
Garde,~S.; Hummer,~G.; Garcia,~A.~E.; Paulaitis,~M.~E.; Pratt,~L.~R. Origin of
  Entropy Convergence in Hydrophobic Hydration and Protein Folding. \emph{Phys.
  Rev. Lett.} \textbf{1996}, \emph{77}, 4966--4968\relax
\mciteBstWouldAddEndPuncttrue
\mciteSetBstMidEndSepPunct{\mcitedefaultmidpunct}
{\mcitedefaultendpunct}{\mcitedefaultseppunct}\relax
\EndOfBibitem
\bibitem[Hummer \latin{et~al.}(1998)Hummer, Garde, Garcia, Paulaitis, and
  Pratt]{lrp:jpcb98}
Hummer,~G.; Garde,~S.; Garcia,~A.~E.; Paulaitis,~M.~E.; Pratt,~L.~R.
  Hydrophobic Effects on a Molecular Scale. \emph{J. Phys. Chem. B}
  \textbf{1998}, \emph{102}, 10469--10482\relax
\mciteBstWouldAddEndPuncttrue
\mciteSetBstMidEndSepPunct{\mcitedefaultmidpunct}
{\mcitedefaultendpunct}{\mcitedefaultseppunct}\relax
\EndOfBibitem
\bibitem[Lebowitz and Percus(1961)Lebowitz, and Percus]{lebowitz:61a}
Lebowitz,~J.~L.; Percus,~J.~K. Long-Range Correlations in a Closed System with
  Applications to Nonuniform Fluids. \emph{Phys. Rev.} \textbf{1961},
  \emph{122}, 1675--1691\relax
\mciteBstWouldAddEndPuncttrue
\mciteSetBstMidEndSepPunct{\mcitedefaultmidpunct}
{\mcitedefaultendpunct}{\mcitedefaultseppunct}\relax
\EndOfBibitem
\bibitem[Lebowitz and Percus(1961)Lebowitz, and Percus]{lebowitz:61b}
Lebowitz,~J.~L.; Percus,~J.~K. Thermodynamic Properties of Small Systems.
  \emph{Phys. Rev.} \textbf{1961}, \emph{124}, 1673--1681\relax
\mciteBstWouldAddEndPuncttrue
\mciteSetBstMidEndSepPunct{\mcitedefaultmidpunct}
{\mcitedefaultendpunct}{\mcitedefaultseppunct}\relax
\EndOfBibitem
\bibitem[Rom\'an \latin{et~al.}(1997)Rom\'an, White, and Velasco]{velasco:97}
Rom\'an,~F.~L.; White,~J.~A.; Velasco,~S. Fluctuations in an Equilibrium
  Hard-Disk Fluid: {Explicit Size Effects}. \emph{J. Chem. Phys.}
  \textbf{1997}, \emph{107}, 4635--4641\relax
\mciteBstWouldAddEndPuncttrue
\mciteSetBstMidEndSepPunct{\mcitedefaultmidpunct}
{\mcitedefaultendpunct}{\mcitedefaultseppunct}\relax
\EndOfBibitem
\bibitem[{El Hage} \latin{et~al.}(2018){El Hage}, H\'edin, Gupta, Meuwly, and
  Karplus]{Meuwly:2018}
{El Hage},~K.; H\'edin,~F.; Gupta,~P.~K.; Meuwly,~M.; Karplus,~M. Valid
  Molecular Dynamics Simulations of Human Hemoglobin Require a Surprisingly
  Large Box Size. \emph{eLife} \textbf{2018}, \emph{7:e35560}\relax
\mciteBstWouldAddEndPuncttrue
\mciteSetBstMidEndSepPunct{\mcitedefaultmidpunct}
{\mcitedefaultendpunct}{\mcitedefaultseppunct}\relax
\EndOfBibitem
\bibitem[Kale \latin{et~al.}(1999)Kale, Skeel, Bhandarkar, Brunner, Gursoy,
  Krawetz, Phillips, Shinozaki, Varadarajan, and Schulten]{namd}
Kale,~L.; Skeel,~R.; Bhandarkar,~M.; Brunner,~R.; Gursoy,~A.; Krawetz,~N.;
  Phillips,~J.; Shinozaki,~A.; Varadarajan,~K.; Schulten,~K. NAMD2: Greater
  Scalability for Parallel Molecular Dynamics. \emph{J. Comput. Phys.}
  \textbf{1999}, \emph{151}, 283\relax
\mciteBstWouldAddEndPuncttrue
\mciteSetBstMidEndSepPunct{\mcitedefaultmidpunct}
{\mcitedefaultendpunct}{\mcitedefaultseppunct}\relax
\EndOfBibitem
\bibitem[Plimpton(1995)]{plimpton:jcop1995}
Plimpton,~S. Fast Parallel Algorithms for Short-Range Molecular Dynamics.
  \emph{J. Comput. Phys.} \textbf{1995}, \emph{117}, 1--19\relax
\mciteBstWouldAddEndPuncttrue
\mciteSetBstMidEndSepPunct{\mcitedefaultmidpunct}
{\mcitedefaultendpunct}{\mcitedefaultseppunct}\relax
\EndOfBibitem
\bibitem[Rackers \latin{et~al.}(2018)Rackers, Wang, Lu, Laury, Lagardere,
  Schnieders, Piquemal, Ren, and Ponder]{tinker8}
Rackers,~J.~A.; Wang,~Z.; Lu,~C.; Laury,~M.~L.; Lagardere,~L.;
  Schnieders,~M.~J.; Piquemal,~J.-P.; Ren,~P.; Ponder,~J.~W. Tinker 8: Software
  Tools for Molecular Design. \emph{J. Chem. Theory Comput.} \textbf{2018},
  \emph{14}, 5273--5289\relax
\mciteBstWouldAddEndPuncttrue
\mciteSetBstMidEndSepPunct{\mcitedefaultmidpunct}
{\mcitedefaultendpunct}{\mcitedefaultseppunct}\relax
\EndOfBibitem
\bibitem[VandeVondele \latin{et~al.}(2005)VandeVondele, Krack, Mohamed,
  Parrinello, Chassaing, and Hutter]{cp2knew}
VandeVondele,~J.; Krack,~M.; Mohamed,~F.; Parrinello,~M.; Chassaing,~T.;
  Hutter,~J. {QUICKSTEP}: {Fast} and Accurate Density Functional Calculations
  Using a Mixed {Gaussian} and Plane Waves Approach. \emph{Comp. Phys. Comm.}
  \textbf{2005}, \emph{167}, 103--128\relax
\mciteBstWouldAddEndPuncttrue
\mciteSetBstMidEndSepPunct{\mcitedefaultmidpunct}
{\mcitedefaultendpunct}{\mcitedefaultseppunct}\relax
\EndOfBibitem
\end{mcitethebibliography}

\providecommand{\latin}[1]{#1}
\makeatletter
\providecommand{\doi}
  {\begingroup\let\do\@makeother\dospecials
  \catcode`\{=1 \catcode`\}=2 \doi@aux}
\providecommand{\doi@aux}[1]{\endgroup\texttt{#1}}
\makeatother
\providecommand*\mcitethebibliography{\thebibliography}
\csname @ifundefined\endcsname{endmcitethebibliography}
  {\let\endmcitethebibliography\endthebibliography}{}

\end{document}